\documentstyle[epsfig]{article}

 \makeatletter
 \def\section{\@startsection {section}{1}{\z@}{3.5ex plus 1ex minus
    .2ex}{2.3ex plus .2ex}{\sc }}
 \def\subsection{\@startsection{subsection}{2}{\z@}{3.25ex plus 1ex minus
   .2ex}{1.5ex plus .2ex}{\small \sc }}
 \def\subsubsection{\@startsection{subsubsection}{2}{\z@}{3.25ex plus 1ex minus
   .2ex}{1.5ex plus .2ex}{\small \sc }}
  \def\appendix{\par\clearpage
  \setcounter{section}{0}
  \setcounter{subsection}{0}
  \@addtoreset{equation}{section}
  \def\@sectname{Appendix~}
  \def\theequation{\thesection.\arabic{equation}}
  \def\thesection{\Alph{section}}}
\makeatother

\newcommand{\be}{\begin{equation}}
\newcommand{\ee}{\end{equation}}
\newcommand{\bea}{\begin{eqnarray}}
\newcommand{\eea}{\end{eqnarray}}
\newcommand{\noi}{\noindent}
\newcommand{\pss}{\protect\scriptscriptstyle}
\newcommand{\pst}{\protect\textstyle\scriptscriptstyle}
\newcommand{\hq}{\hat{q}}
\newcommand{\mq}{\left|\hq\right|}

\newcommand{\sq}{\mq^2\,\delta_{\mu\nu}\:-\:\hq_\mu\,\hq_\nu^*}
\newcommand{\epf}{{\cal E}^{\pss (1)}}
\newcommand{\eps}{{\cal E}^{\pss (2)}}
\newcommand{\tq}{\tilde{q}}
\newcommand{\tqp}{\tilde{q}'}
\newcommand{\tqpp}{\tilde{q}''}
\newcommand{\tqm}{\tilde{q}/2}
\newcommand{\tqpm}{\tilde{q}'/2}
\newcommand{\tqppm}{\tilde{q}''/2}

\newcommand{\f}{\frac}

\begin{document}
\thispagestyle{empty}
\parskip=12pt
\raggedbottom

\def\mytoday#1{{ } \ifcase\month \or
 January\or February\or March\or April\or May\or June\or
 July\or August\or September\or October\or November\or December\fi
 \space \number\year}
\noindent
\hspace*{9cm} BUTP-97/24\\
\vspace*{1cm}
\begin{center}
{\LARGE The fixed point action for the Schwinger model: \\
a perturbative approach}\footnote{Work supported by Fondazione 
``A. Della Riccia'' (Italy) and Ministerio de Educaci\'on y 
Cultura (Spain).}

\vspace{1cm}

Federico Farchioni and Victor Laliena\\
Institute for Theoretical Physics \\
University of Bern \\
Sidlerstrasse 5, CH-3012 Bern, Switzerland

\vspace{0.5cm}

\mytoday \\ \vspace*{0.5cm}

\nopagebreak[4]

\begin{abstract}
We compute the fixed point action of a properly defined
renormalization group transformation for the Schwinger model
through an expansion in the gauge field.  
It is local, with couplings exponentially suppressed with the distance. 
We check its perfection 
by computing the 1-loop mass gap at finite spatial volume,
finding only exponentially vanishing cut off effects, in contrast with the
standard action, which is affected by large power-like cut off
effects. We point out that the 1-loop mass gap calculation
provides a check of the classical perfection of the fixed point action,
and not of the 1-loop perfection, as could be naively expected.
\end{abstract}

\end{center}
\eject

\section{\bf Introduction.}

The discretization of the space-time into a lattice 
provides a non-perturbative regularization 
of a quantum field theory which, in addition, allows numerical
simulations. The lattice spacing is finite in any Monte Carlo simulation,
and the distortions on the physical quantities induced 
by the discretization  (cut off effects) strongly restrict 
the accuracy of the method.
The naive procedure - consisting in approaching the continuum by progressively 
reducing the lattice spacing - has to cope soon with the problem of 
the divergence of computational time and memory space. 
Therefore, new methods have been studied 
in order to reduce the lattice artifacts at their origin, i.e. 
at the level of the lattice action.

The first method, due to Symanzik, consists in adding to the simplest 
discretized action (standard action) higher order operators 
which cancel the cut off effects 
to a given order in the lattice spacing (usually the leading one, 
$O(a^2)$ for bosonic theories 
and $O(a)$ in presence of fermions) and in the coupling constant; 
this method is designed for perturbation theory.

The second method uses the Wilson Renormalization Group (RG) theory, 
and is inherently non-perturbative. 
In this case, one can be very ambitious and
search for perfect actions~\cite{hn0}, which, by definition, reproduce 
exactly the continuum independently of the value of the lattice spacing. 
The existence of perfect actions follows from the existence 
of a renormalized trajectory in the space
of parameters of the theory: any action located on the renormalized 
trajectory is a perfect action~\cite{hn0}.   

A more modest and realistic goal is the determination of a classically 
perfect action, i.e. an action which - in principle - eliminates 
the cut off effects with restriction to 
the classical properties of the theory. This action is related~\cite{hn0} to 
the fixed point (FP) - lying on the critical surface - of a given 
RG transformation.
In the case of a theory which attains the continuum for weak couplings,
the FP problem is reduced to the solution of a saddle point equation.
Although not perfect, the FP action represents a huge step toward 
the elimination of the cut off effects in comparison with 
naive discretizations, and a considerable improvement is observed even 
with respect to the Symanzik 
improved actions~\cite{hn0,hnymn,aless}.

In view of an application of these ideas to the numerical solution of
lattice QCD, the fermion sector must be well understood. FP actions
for gauge theories with interacting fermions have not been 
extensively studied yet. 
In this paper we will study the fermion-gauge field FP interactions in a 
much simpler case than QCD, the Schwinger model, which belongs to the class of
theories for which the computation of a FP action is a classical
saddle point problem. Since its gauge group is abelian, it is 
possible to formulate the lattice regularization with non-compact 
gauge fields, and to solve analytically the pure gauge sector.
Therefore, we are able to concentrate the numerical effort in the 
fermion problem. We will test the perfection of the FP action by 
computing the 1-loop mass gap in a finite volume - 
a circle of length $L$ -  using the standard action as a ``control'' action.

The remaining of the paper is organized as follows. 
In Section~\ref{sec:fpa} we review briefly
the formalism of the FP actions. In Section~\ref{sec:schw}  
we recall some features of
the Schwinger model which are relevant for this work. Section~\ref{sec:gau} is 
devoted to the study of the FP action for the pure gauge sector. 
In Section~\ref{sec:ferm} we address ourselves to the computation 
of the fermion part of the FP action; the problem is treated perturbatively - 
i.e. in an expansion in the gauge field. In Section~\ref{sec:mass} 
we check the perfection of the FP action, providing  the computation of the 
mass gap to 1-loop. For comparison, we compute the mass gap 
also for the standard action. We end the paper with a discussion about 
the 1-loop perfection of the FP action (Section~\ref{sec:perf})
and the conclusions (Section~\ref{sec:conc}).      

\section{\bf Fixed point actions.}
\label{sec:fpa}

In this Section we briefly describe the general method for 
the computation of the FP action in the case of theories reaching the
continuum for small couplings, referring to the 
literature~\cite{hnymp,wnf,kunszt} for the details. 
For the sake of definiteness, we restrict the notation to case of the 
$U(1)$ gauge group in the non-compact formulation.

A general form of the action of a lattice-regularized gauge theory is 
\be
S\;=\;\beta\,S_g(A)\:+\:\bar\psi\,\Delta(U)\,\psi\, ,
\ee
\noi
where $A$ denotes the gauge field configuration; $\Delta$ is a suitable
fermion matrix which depends on the gauge field through the link variable
$U_{\mu}=\exp(iA_{\mu})$. The detailed form of the action is here 
not relevant, apart from the requirement of gauge invariance 
and recovering of the classical continuum limit. In the fermion sector some
care is necessary in order to ensure that the doublers decouple in the
continuum limit. The partition function is given by the path integral
\be
{\cal Z}\;=\;\int\,\left[dA\,d\bar\psi\,d\psi\right]\,\exp\,\left[\,
-\,\beta\,S_g(A)\:-\:\bar\psi\,\Delta(U)\,\psi\,\right]\, .
\ee
\noi

A RG transformation can be defined, which maps the action $S$ 
onto $S^\prime$, the latter action depending on fields defined 
on a coarser lattice:
\bea
&&\hspace*{-1 cm}
\exp\,\left[\,-\,\beta^\prime\,S_g^\prime(A^\prime)\:-\:
S_F^\prime(\bar\psi^\prime,\psi^\prime,U^\prime)\,\right]\;= \nonumber \\
& &\hspace*{1 cm} \int\,\left[dA\,d\bar\psi\,d\psi\right]\,\exp\,\left[
-\,\beta\,S_g(A)\:-\:\bar\psi\,\Delta(U)\,\psi\:-\:p\,{\cal K}_g(A^\prime,A) \right.
 \nonumber \\
& &\hspace*{1 cm}
\left. -\:\kappa_F\,{\cal K}_F(\bar\psi^\prime,\psi^\prime,\bar\psi,\psi,U)
\,\right]\, , \label{rgtdef}
\eea
\noi
where the primed fields are the degrees of freedom on the coarser lattice,
defined through the gauge invariant kernels \ ${\cal K}_g$ \ and \ 
${\cal K}_F$.
The parameters $p$ and $\kappa_F$ can be chosen arbitrarily\footnote{The
choice is somehow restricted  by the request that the RG transformation 
converges to a FP when iterated infinitely many times.}; in particular,
we can take $p=\beta\,\kappa_g$. Note that in general the fermion action is 
not quadratic in the fermion fields after a RG transformation.

In the class of theories under interest (including asymptotically 
free non-abelian gauge theories and the Schwinger model), 
the critical surface (where
the continuum is attained) is at $\beta=\infty$. 
The iteration of the RG transformation starting on this surface
converges to a FP. 
When $\beta\rightarrow\infty$ the integral on the gauge 
degrees of freedom in the r.h.s. of Eq.~(\ref{rgtdef}) is saturated by the 
saddle point configuration; the solution of the recursion is then:
\bea
& & S_g^\prime(A^\prime)\;=\;\min_{\{ A \} }\,\left[S_g(A)\:+\:
\kappa_g\,{\cal K}_g (A^\prime,A)\,\right] \, ,  \label{sapo} \\
& & S_F^\prime (\bar\psi^\prime,\psi^\prime,U^\prime)\;=\;-\ln\,
\int\,\left[d\bar\psi\,d\psi\right]\,\exp\,\left[\,
-\:\bar\psi\,\Delta(U(A^{\pst min}))\,\psi \right. \nonumber \\
& &\left.\;\;\;\;\;\;\;\;\;\;\;\;\;\;\;\;\;\;\;\;\;\;\;\;\;\;\;\;\;\;\;\;\;\;\;\;
-\:\kappa_F\,{\cal K}_F(\bar\psi^\prime,\psi^\prime,
\bar\psi,\psi,U(A^{\pst min}))\,\right]\, , \label{sapof}
\eea
\noi
where $A^{\pst min}$ is the fine gauge field configuration which minimizes 
the r.h.s. of Eq.~(\ref{sapo}), depending on the coarse 
configuration $A^\prime$; of course: $\beta^\prime=\beta=\infty$.
In this limit the problem is equivalent to a classical minimization problem
plus a Grassmann integration.

If the fermion kernel is quadratic in the fermion fields,
the fermion action at $\beta=\infty$ remains quadratic after a RG  
transformation, as is evident from Eq.~(\ref{sapof}). 
The FP action is defined as:
\be
S^{\pss FP}\;=\;\beta\,S^{\pss FP}_g(A)\:+\:\bar\psi\,\Delta^{\pss FP}(U)\,\psi\, ,
\ee
\noi
where $S^{\pss FP}_g$ and $\Delta^{\pss FP}$ are the self-reproducing 
solutions of Eq.~(\ref{sapo}) and  Eq.~(\ref{sapof}) respectively.
The FP action is then bilinear in the fermion fields; 
this is a very important issue, mainly for what concerns numerical simulations.

\section{\bf General considerations about the Schwinger model.}
\label{sec:schw}

\subsection{The continuum Schwinger model.}

In this Section we shall review some features of the Schwinger model  
which are relevant for the following. The Schwinger model is the 
transcription to $1+1$ dimensions of the usual $3+1$ QED~\cite{schw}. Its
euclidean lagrangian reads (for the massive model): 
\be
{\cal L}\;=\;\frac{1}{4}\,\sum_{\mu\nu}F_{\mu\nu}(x)F_{\mu\nu}(x)\:
+\:\bar\psi (x)\,\left[\,\sum_\mu\gamma_\mu\partial_\mu\:+\:
i\,e\,\sum_\mu\gamma_\mu\, A_\mu (x)\:+\:m_f\,\right]\,\psi (x)\, .
\ee
\noi
We use the simplest representation of the euclidean Dirac matrices in $1+1$
dimensions, given by the Pauli matrices:
\be
\gamma_0\;=\;\sigma_1\, ,\;\;\;\;\;\;\;\;\gamma_1\;=\;\sigma_2\, ,
\;\;\;\;\;\;\;\; 
\gamma_5\;=\;i\,\gamma_0\,\gamma_1\;=\;-\,\sigma_3 \, . \label{diracmat}
\ee

The physical spectrum of the massless ($m_f=0$) single-flavor model 
contains only a free boson of mass $e/\sqrt{\pi}$~\cite{schw}. 
This boson couples to the gauge field, allowing the computation of 
its mass from the gauge propagator, which can be written as
\be
G_{\mu\nu}(q)\;=\;\frac{1}{q^2-q^2\,\Pi(q^2)}\,
\left(\,\delta_{\mu\nu}\:-\:\frac{q_\mu q_\nu}{q^2}\,\right)\:+\:
\frac{1}{\xi\, q^2}\,\frac{q_\mu q_\nu}{q^2} \, ,
\ee
\noi
where $\Pi(q^2)$ is defined through the vacuum polarization 
tensor $\Pi_{\mu\nu}(q)$:
\be 
\Pi_{\mu\nu}(q)\;=\;\left(\,q^2\,\delta_{\mu\nu}\:-\:q_\mu q_\nu\,\right)\,
\Pi(q^2) \, .\label{cvacpol}
\ee

\begin{figure}[t]
\centering
\mbox{\psfig{file=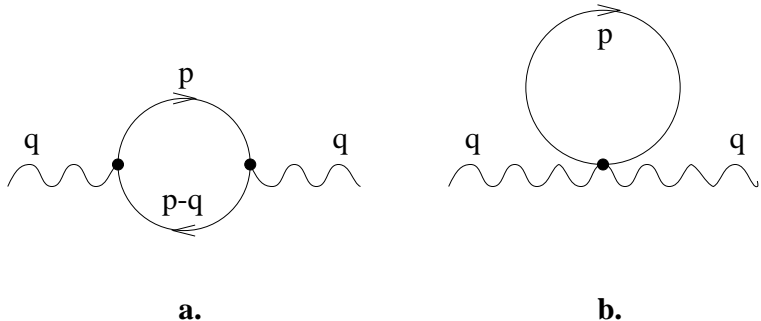,width=10 truecm, angle=0}}
\caption{Diagrams contributing to the 1-loop vacuum polarization.}
\label{fig:fey}
\vspace{0.5cm}
\end{figure}

The theory is solved at 1-loop in perturbation theory,
since fermion loops with more than two photons vanish~\cite{bk}:
only in this way both the vector and chiral Ward identities 
can be satisfied\footnote{Since the integral associated  
with a fermion loop with  more than two photons is both UV and IR
convergent, there is no possibility of violating any of the Ward identities
through an anomaly.}.
The fermion loop with two photons (Fig.~\ref{fig:fey}-a) is
superficially UV divergent and asks for a regularization. Using for example the
Pauli-Villars procedure, the chiral symmetry is explicitly broken,
and it is not restored when removing the cut off; the previous argument 
is therefore evaded, and an UV finite but non-zero result is obtained. 
As a consequence, the diagram gives the exact vacuum polarization:
\be 
\Pi(q^2)\;=\;-\frac{e^2}{\pi}\,\frac{1}{q^2}\, . \label{csvacpol}
\ee

In the following we will be interested in the formulation of the model on
a spatial circle of length $L$. In the continuum, this model has been studied 
for the first time
in~\cite{manton}: its physical content turns out to be the same as 
in the $L=\infty$ case, which is not surprising taking into 
account~\cite{manton} that the model can be bosonized into a free-field theory.

\subsection{Lattice regularization.}

The simplest action for the model regularized on the lattice, with non-compact
gauge fields and Wilson fermions, is:
\be
S\;=\;\frac{\beta}{2}\,\sum_{x\, ,\,\mu<\nu}\,F^2_{\mu\nu}(x)\:+\:
\sum_{x,x^\prime}\,\bar{\psi}_x\,\Delta(x,x^\prime\, ;\,U)\,\psi_{x^\prime}\, ,
\label{wac}
\ee
\noi
where $\beta=\frac{1}{g^2}$ and $g$ is the dimensionless lattice coupling
constant, which is related to the dimensionful electric charge $e$ 
through $g=ea$ ($a$ is the lattice spacing). The lattice strength field 
tensor $F_{\mu\nu}(x)$ is given by:
\be
F_{\mu\nu}(x)\;=\;A_\mu(x)\:+\:A_\nu(x+\hat\mu)\:-\:A_\mu(x+\hat\nu)\:-\:
A_\nu(x)\, ,
\ee
\noi
and the fermion matrix takes the form
\begin{displaymath}
\Delta(x,x^\prime\, ;\,U) \;=\; (m_f+4)\,\delta_{x\,x^\prime}
\end{displaymath}
\be
-\:\frac{1}{2}
\sum_\mu\,\left[\,\left(1-\gamma_\mu\right)\,U_\mu(x)\,\delta_{x^\prime\, ,\,
x+\hat\mu}\:+\:\left(1+\gamma_\mu\right)\,U_\mu^\dagger(x^\prime)\,
\delta_{x^\prime\,,\,x-\hat\mu}\,\right]\, .
\ee
\noi
The coupling between the fermions and the gauge field must be compact in order
to ensure gauge invariance.
We will work with massless fermions, $m_f=0$.

The gauge field $A$ is normalized in such a way that, when expanding the
fermion action in powers of $A$:
\be
\bar\psi\,\Delta(A)\,\psi \;=\;\sum_{x\,y}\bar\psi_x\,D(x-y)\,\psi_y 
\:+\: i\sum_{x\,y\,r}\,\sum_\mu
A_\mu(r)\,\bar\psi_x\,R^{\pss (1)}_\mu(x-r\, ,\,y-r)\,\psi_y\:+\:\ldots\, ,
\ee
  \noi
the first order vertex $R^{\pss (1)}_\mu$ at zero momentum equals the
continuum vertex with unit charge:
\be
\sum_{x,y}\,R^{\pss (1)}_\mu(x,y)\;=\;\gamma_\mu \, . \label{normal}
\ee
\noi

With the non-compact formulation for the pure gauge sector, a gauge fixing
term is necessary in order to have a well defined path integral.
The free gauge propagator can be written as
\be
G_{\mu\nu}(q)\;=\;\frac{1}{\mq^2}\,
\left(\,\delta_{\mu\nu}\:-\:\frac{\hq_\mu\,\hq^*_\nu}{\mq^2}\,\right)\:+\:
\frac{1}{\xi}\,\frac{\hq_\mu\,\hq^*_\nu}{\mq^4}
\, , \label{latprop0}
\ee
\noi
where $\xi$ is the gauge fixing parameter,
$\hq_\mu = \exp(iq_\mu) - 1$,\ $\mq^2 = \sum_\mu\hq_\mu^*\hq_\mu$ and
$q=(q_0,q_1)$.
The full inverse propagator can be expressed as
\be
G^{{\pst full}\,-1}_{\mu\nu}(q)\;=\;G_{\mu\nu}^{-1}(q)\:-\:
\Pi_{\mu\nu}(q)\, , \label{igp}
\ee
\noi
where $\Pi_{\mu\nu}(q)$ is the lattice vacuum polarization tensor. Gauge
invariance implies the  Ward identity, which on the lattice reads:
\be
\sum_\mu\,\hq^*_\mu\,\Pi_{\mu\nu}(q)\;=\;0\, . \label{piwi}
\ee

\subsection{Mass gap and cut off effects.}
\label{sec:cuteff}

The mass of a particle coupled to the gauge field is obtained by 
looking at the zeros of the eigenvalues of the inverse gauge propagator 
(\ref{igp}) for zero spatial momentum. 
From the Ward identity we have
\be
\Pi_{00}(q_0,q_1=0)\;=\;\Pi_{01}(q_0,q_1=0)\;=\;
\Pi_{10}(q_0,q_1=0)\;=\;0\, ,
\ee
\noi
and therefore the equation for the mass gap is
\be
\left|\hq_0\right|^2\:-\:\Pi_{11}(q_0,q_1=0)\;=\;0 \, . \label{masseq}
\ee
\noi
The solution of this equation is purely imaginary, at least when the lattice
spacing is small enough,
and depends on  $L$,\ $a$ and $g$:\  $q_0=i\, m(g,a/L)$. The quantity 
 $m$ is the dimensionless lattice mass, which is related to the 
physical mass through the relation $m_{\pss ph}=m/a$.   

It is possible to compute the lattice vacuum polarization perturbatively
through an expansion in powers of the lattice coupling constant $g$: 
\be 
\Pi_{\mu\nu}(q\,,a/L)\;=\;g^2\:
\Pi_{\mu\nu}^{\pss (1)}(q,\,a/L)\;+\;
g^4\:\Pi_{\mu\nu}^{\pss (2)}(q,\,a/L)\;+\;\ldots\, , \label{pvp}
\ee
 \noi
where $q$ is the (dimensionless) lattice momentum and we have explicitly
written the dependence in $a/L$.
On the lattice the contribution of the higher order loops is non-zero,
but it vanishes in the continuum limit; so it constitutes a pure cut off
effect. The 1-loop vacuum polarization is given by the two diagrams of
Fig.~\ref{fig:fey}.

In perturbation theory the mass gap is computed order by order in $g$:
\be
m(g,\,a/L)\;=\;g\:m^{\pss (1)}(a/L)\;+\;
g^3\:m^{\pss (2)}(a/L)\;+\;\ldots\, . \label{pmass}
\ee
\noi
Combining (\ref{masseq}), (\ref{pvp}) and (\ref{pmass}) we get
\be
m^{\pss (1)}(a/L)\;=\;\lim_{q_0\rightarrow0}\sqrt{-\Pi_{11}^{\pss (1)}(q_0,q_1=0,\,a/L)}\, .
\label{limit}
\ee
\noi
The scaling limit $a\rightarrow 0$ gives the continuum mass 
$m_{\pss ph}^{\pss (c)}=\frac{e}{\sqrt{\pi}}$ in the following way:
\be
\frac{1}{e}\,m_{\pss ph}(ea,\,a/L)\;=\;m^{\pss (1)}(a/L)\:+\:
e^2\,a^2\,m^{\pss (2)}(a/L)\;+\;\ldots\;\rightarrow\;
\frac{m_{\pss ph}^{\pss (c)}}{e}\;=\;\frac{1}{\sqrt\pi}\, .
\label{cont}
\ee
From the above formula is explicit that higher order corrections to the mass
are pure cut off effects.
\noi
In general we can write:
\be
m^{\pss (1)}(a/L)\;=\;\frac{1}{\sqrt\pi}\:+\:F^{\pss (1)}(a/L)\, ,
\label{contino}
\ee
\noi
Eq.~(\ref{cont}) implies $F^{\pss (1)}(0)\,=\,0$. 

These general considerations, based on dimensional analysis and on the UV
finiteness of the model, imply that in the $L=\infty$ case 
no cut off effects appear in the mass gap at 1-loop order, independently
of the lattice action chosen. However, they can be present as soon the model  
is put on a circle of finite length. These cut off effects depend on the 
form of the lattice action, and, as discussed in the Introduction, 
they can be strongly suppressed, and even removed, by choosing a proper 
lattice action. It is the aim of this work to see to what extent 
the cut off effects are suppressed in the 1-loop calculations, 
when using a FP action, which is in principle perfect 
only at the classical level, i.e. at the tree level.

In the rest of the paper we will describe the construction of a FP action for
the Schwinger model and we will discuss its cut off effects in the mass gap,
comparing them with those of the standard action (\ref{wac}).
The first step in the construction of the whole FP action is the determination
of the FP action for the pure gauge sector, using the saddle point equation 
(Eq.~(\ref{sapo})), and of the minimizing configuration $A^{\pst min}$ 
as a function of the coarse configuration $A'$.

\section{\bf FP action for the pure gauge sector.}
\label{sec:gau}

The quantum theory of the free electromagnetic field in $1+1$ dimensions is
equivalent to the quantum mechanics of a rotor~\cite{manton}, for which it has
been shown that ultralocal perfect and FP actions~\cite{wrotor} can be 
invented.
The existence of ultra-local FP actions for two-dimensional abelian 
gauge field theories had been already pointed out in~\cite{wnf}. 
In this Section we will define a block transformation for the gauge 
field which leads to the standard non-compact action as FP action. 

We use the techniques and notations of~\cite{hnymp}, and we refer the reader
to this paper for further details.  
 
\subsection{RG transformation and fixed point action.}

Let us start with a standard non-compact lattice action,  for example the 
Wilson action (\ref{wac}).
The pure gauge part describes a free field. Its action in momentum space is
\be
S_g\;=\;\frac{1}{2}\,\frac{1}{V}\,\sum_q\,A_\mu^*(q)\,\left(\,\sq\:+\:
\xi\,\hq_\mu\,\hq_\nu^*\,\right)\,A_\nu(q) \, .
\ee
\noi
We have added to the action a temporary gauge fixing term; it will 
be removed at the end of the calculation. 

We choose a simple gauge kernel:
\be
{\cal K}_g(A^\prime,A)\;=\;\sum_{x_B,\mu}\,\left[\,A^\prime_\mu(x_B)\:-\:
\frac{1}{2}\left(\,A_\mu(2x_B)\,+\,A_\mu(2x_B+\hat\mu)\,\right)\,\right]^2\, ,
\label{gaugekernel}
\ee
\noi
which is of course gauge invariant, so ensuring the conservation of the 
gauge invariance under RG transformations.
The variables $x_B$ label the sites on the coarse lattice in units of 
its doubled lattice spacing: the site $x_B$ corresponds to the site
$2x_B$ in the units of the original lattice. Since the pure gauge
sector - including the kernel - is gaussian, the minimization
problem is equivalent to the solution of the exact gaussian integration.
In Appendix~\ref{app:a} it is shown that the FP propagator is given by
\be
G_{\mu\nu}^{\pss FP}(q)\;=\;\left(\,f_\mu(q)\:+\:\frac{2}{\kappa_g}\,\right)\,
\delta_{\mu\nu}\;+\;(\xi^{-1}\,-\,1)\,g(q)\,\hq_\mu\,\hq_\nu^* \, ,
\label{fpgapro}
\ee
\noi
where the functions $f_\mu$ and $g$ are given in the formula (\ref{apfgeq}) of
the same Appendix.

The inverse propagator $\rho_{\mu\nu}^{\pss FP}(q)$ has a well defined limit 
when $\xi\rightarrow 0$, which defines a gauge invariant action. 
Using the relation 
\be
\left|\,\hq_0\,\right|^2\,f_1(q)\:+\:\left|\,\hq_1\,\right|^2\,f_0(q)\;=\;1\;,
\ee
\noi
we find:
\be
\rho_{\mu\nu}^{\pss FP}(q)\;=\;\frac{1}{1+\frac{2}{\kappa_g}\mq^2}\,
\left(\,\sq\,\right)\, .
\ee
\noi
Taking $\kappa_g\rightarrow\infty$ we recover the original standard 
non-compact action as FP action. It is ultralocal, involving only 
nearest-neighbors interactions.

We need also the minimizing gauge field as a function of the coarse field
$A^\prime$. Since the problem is quadratic, the relation is 
linear~\cite{hnymp} : 
\be
A_\mu^{min}(\frac{q+2\pi\,l}{2})\;=\;\sum_\nu\,Z_{\mu\nu}(\frac{q+2\pi\,l}{2})
\,A_\nu^\prime(q)\, , \label{minfi}
\ee
\noi
where $l_\mu=0,1$. In Appendix~\ref{app:a2} can be found 
the explicit expression for
$Z_{\mu\nu}$ and for other related functions useful for the perturbative
determination of the fermion-photon FP interaction.

\subsection{Discussion.}

At first sight, it might seem surprising that the standard action, 
which couples only first neighbors, can be  the FP of some 
RG transformation. Indeed, as already pointed out, FP actions 
are believed to be classically perfect. 
In the present case, in particular, this nearest neighbor action
is expected to reproduce the classical properties of the two-dimensional
electrodynamics.
In $1+1$ dimensions, however, no photon is described by the Maxwell field. 
In fact, gauge symmetry constrains the dynamics so strongly that only 
a non-propagating Coulomb field can exist. The best way to see
this is to consider the propagator in the Coulomb gauge. 
In this gauge, the ``photon'' propagator is given by
\be
G^{\pss c}_{00}(q)\;=\;\frac{1}{q_1^2}\, ,\;\;\;\;\;\;\;\;G^{\pss c}_{01}(q)
\;=\;G^{\pss c}_{10}(q)\;=\;G^{\pss c}_{11}(q)\;=\;0\, .
\ee
\noi
We see that there is no propagating degree of freedom and therefore 
no spectrum. Taking the Fourier transform - subtracting the propagator at 
zero spatial distance to avoid the IR  divergence - we find the 
Coulomb potential in two dimensions:
\be
G^{\pss c\, (R)}_{00}(x)\;=\;G^{\pss c}_{00}(x)\:-\:G^{\pss c}_{00}(x_0,x_1=0)
\;=\;-\,\delta(x_0)\:\frac{\left|\, x_1\,\right|}{2} \, .
\ee

On the lattice, the Coulomb gauge is fixed by introducing the following gauge
fixing term:
\be
\frac{\xi}{2}\,\sum_x\,\left[\,A_1(x+\hat{1})\,-\,A_1(x)\,\right]^2\, ,
\ee
\noi 
and taking the $\xi\rightarrow\infty$ limit. The lattice propagator is
\be
G^{\pss c}_{00}(q)\;=\;\frac{1}{\left|\hq_1\right|^2}\, ,\;\;\;\;\;\;\;
G^{\pss c}_{01}(q)\;=\;G^{\pss c}_{10}(q)\;=\;G^{\pss c}_{11}(q)\;=\;0\, .
\ee
\noi
Again, there is no propagating degree of freedom. If we go to coordinate
space, renormalizing the IR divergence, we obtain
\be
G^{\pss c\, (R)}_{00}(n)\;=\;G^{\pss c}_{00}(n)\:-\:
G^{\pss c}_{00}(n_0,n_1=0)
\;=\;-\,\delta_{n_0,0}\:\frac{\left|\,n_1\,\right|}{2}\, .
\label{lcp}
\ee
\noi
This is the perfect Coulomb potential on the lattice, which is the Green
function of the perfect laplacian in one dimension. It has been obtained
from the standard nearest-neighbors non-compact lattice action. This is a
check of the classical perfection of this action. 

Alternatively, one can compute the potential between 
heavy charges through the Wilson loop\footnote{This is a theoretically
cleaner way to obtain the potential, since it goes through the
calculation of the expectation value of a gauge-invariant quantity.}.
Again, one gets the perfect Coulomb potential 
in $1+1$ dimensions (Eq.~(\ref{lcp})).
A cut off free lattice potential requires a perfect  Wilson
loop operator as well~\cite{hnymp}. 
In this case, the usual definition of the Wilson loop operator 
is already perfect; indeed, as it can be easily checked, 
it goes into itself under the RG transformation 
defined by the kernel of Eq.~(\ref{gaugekernel}) 
if the $\kappa_g\rightarrow\infty$ limit is taken.

\section{\bf The fermion sector.}
\label{sec:ferm}

After the solution of the pure gauge part, we will concentrate our 
effort in the fermion sector. In this case there is no systematic method 
allowing an exact analytic solution, and we will rely
on an expansion in the gauge field $A_{\mu}$.
A non-perturbative study in this same context has been 
carried out in~\cite{graz}.

We start reviewing the free fermion problem, which has been studied
elsewhere (see for example~\cite{wfree,kunszt}). 
In Appendix~\ref{app:b1} we give some details.

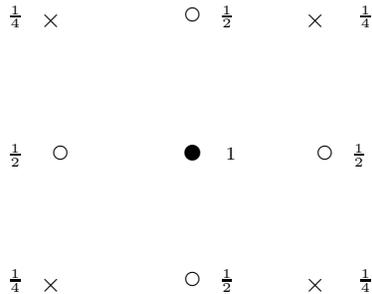
\begin{figure}[t]
\setlength{\unitlength}{0.5 pt}

\begin{center}

\begin{picture}(300,300)

\put (150,150){\circle*{12}}
\put (50,150){\circle{10}}
\put (250,150){\circle{10}}
\put (150,255){\circle{10}}
\put (150,55){\circle{10}}
\put (35,45){$\times$}
\put (35,245){$\times$}
\put (235,45){$\times$}
\put (235,245){$\times$}
\put (170,250){$\scriptstyle\frac{1}{2}$}
\put (170,50){$\scriptstyle\frac{1}{2}$}
\put (175,145){$\scriptstyle 1$}
\put (10,145){$\scriptstyle\frac{1}{2}$}
\put (270,145){$\scriptstyle\frac{1}{2}$}
\put (10,50){$\scriptstyle\frac{1}{4}$}
\put (10,250){$\scriptstyle\frac{1}{4}$}
\put (275,50){$\scriptstyle\frac{1}{4}$}
\put (275,250){$\scriptstyle\frac{1}{4}$}

\end{picture}

\end{center}

\vspace{-1 cm}
\caption{The RG block transformation for free fermions.}
\label{fig:block}
\vspace{0.5cm}
\end{figure}

\subsection{Free fermions.}
\label{subsec:freef}
Let us as first define a block transformation for the fermion fields; 
the form of the quadratic kernel is:
\be
{\cal K}_F\;=\;\sum_{x_B}\,\left[\,\bar\psi^\prime(x_B)\:-\:\bar\Gamma^0(x_B)\,
\right]\,\left[\,\psi^\prime(x_B)\:-\:\Gamma^0(x_B)\,\right] \, ,
\ee
\noi
where 
\be
\Gamma^0(x_B)\;=\;\sum_x\,\omega(2x_B-x)\,\psi_x\;.
\ee

We choose $\Gamma^0(x_B)$ to be proportional to the sum of the fine fields at 
the site $x_B$ and its nearest and next-to-nearest neighbors, each weighted 
with a factor inversely proportional to the number of coarse fields to which 
they are coupled (see Fig.~\ref{fig:block}). Explicitly:
\bea
\Gamma^0(x_B)&=&\frac{\sqrt{2}}{4}\,\left[\,\psi(2x_B)\:+\:
\frac{1}{2}\,\sum_{\lambda=\pm 1}\:\sum_\mu\,\psi(2x_B\,+\,\lambda\hat\mu)
\right. \nonumber \\
&+&\left. \frac{1}{4}\,\sum_{\lambda_0=\pm 1}\:\sum_{\lambda_1=\pm 1}\,
\psi(2x_B\,+\,\lambda_0\hat{0}\,+\,\lambda_1\hat{1})\,\right]\, .
\label{freefk}
\eea
\noi
The global factor $\sqrt{2}/4$ appears by dimensional reasons: in 1+1
dimensions the fermion field has dimension $d_\psi=1/2$; if we take a constant
fine field configuration, the averaged field $\Gamma^0$ must be proportional
to the fine field, with a proportionality factor $2^{d_\psi}$ since we have
increased the lattice spacing by a factor~$2$.

Several block transformations for free fermions were studied 
in~\cite{wfree,kunszt,kunsztalone}. The one considered 
here has at least two good properties: 
it can be managed analytically to obtain a suitable expression for the 
FP propagator and the implementation of gauge invariance 
is straightforward\footnote{In addition it has a faster convergence 
to the FP compared to other non-symmetrical 
definitions~\cite{kunsztalone}.}. 

We write for the FP fermion propagator its most general expression:
\be
D_{\pss FP}(q)\;=\;-\,i\,\sum_\mu\,\gamma_\mu\,\alpha^{\pss FP}_\mu(q)\:+\:
\beta^{\pss FP}(q)\;.
\ee
\noi
The iteration of the RG transformation, starting from the Wilson action,
leads to the following FP propagator:
\bea
& &\alpha^{\pss FP}_\mu(q)\;=\;\sum_{l=-\infty}^{+\infty}\,
\frac{q_\mu\,+\,2\pi\,l_\mu}{(q\,+\,2\pi\,l)^2}\,
\prod_\nu\,\frac{\sin^4\,(q_\nu/2)}
{(q_\nu/2\,+\,\pi\,l_\nu)^4} \label{alphafp} \\
& &\beta^{\pss FP}(q)\;=\;\frac{1}{\kappa_F}\,
\left(\,\frac{328}{217}\:+\:\frac{44}{217}\,(\,\cos q_0\,+\,\cos q_1\,)
\:+\:\frac{18}{217}\,\cos q_0\,\cos q_1\,\right) \;.
\label{betafp} \nonumber \\
& & 
\eea
\noi 
The locality of the FP action is optimal for  $\kappa_F\simeq 4$.
In the subsequent computations we will use this optimized FP action. 

\subsection{Gauge interactions and the perturbative solution.}

In presence of gauge interactions, $\Gamma^0(x_B)$ must be modified 
into a gauge covariant average of the fine fermion fields $\Gamma(x_B;U)$. 
We achieve this in the simplest way, by using the parallel
transport along the simplest symmetric path which joins the 
coarse site $x_B$ to its fine neighbors 
(see Fig.~\ref{fig:blockg}). The explicit expression for $\Gamma(x_B;U)$ 
is reported in Appendix~\ref{app:b2}.

We treat the problem with interaction perturbatively, expanding both the
action and the kernel ${\cal K}_F$ in powers of the weak field $A_{\mu}$
and solving the recursion relation order by order. 

Using translational invariance, the expansion of the fermion action can be
written as follows: 
\bea
S_F &=& \sum_{x\,y}\bar\psi_x\,D^{-1}(x-y)\,\psi_y\:+\:i\sum_{x\,y\,r}\,\sum_\mu
A_\mu(r)\,\bar\psi_x\,R^{\pss (1)}_\mu(x-r\, ,\,y-r)\,\psi_y
\nonumber \\
&+&\sum_{x\,y\,r\,r'}\,\sum_{\mu\nu}\,A_\mu(r)\,A_\nu(r')\,\bar\psi_x\,
R^{\pss (2)}_{\mu\nu}(x-r\, ,\, y-r'\, , r-r')\,\psi_y\:+\:\ldots 
\nonumber \\
\label{sexpan}
\eea
\noi
We call $R^{\pss (1)}_\mu$ the first order vertex and $R^{\pss (2)}_{\mu\nu}$
the second order vertex.

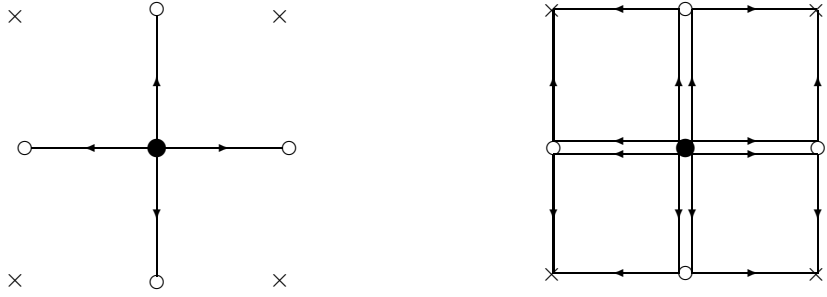
\begin{figure}[t]
\setlength{\unitlength}{0.5 pt}

\begin{center}

\begin{picture}(700,300)

\put (150,150){\circle*{14}}
\put (50,150){\circle{10}}
\put (250,150){\circle{10}}
\put (150,255){\circle{10}}
\put (150,48){\circle{10}}
\put (35,45){$\times$}
\put (35,245){$\times$}
\put (235,45){$\times$}
\put (235,245){$\times$}

\put (150,150){\vector(0,1){55}}
\put (150,205){\line(0,1){45}}
\put (150,150){\vector(0,-1){55}}
\put (150,98){\line(0,-1){45}}
\put (150,150){\vector(1,0){55}}
\put (205,150){\line(1,0){40}}
\put (150,150){\vector(-1,0){55}}
\put (95,150){\line(-1,0){40}}

\put (550,150){\circle*{14}}
\put (450,150){\circle{10}}
\put (650,150){\circle{10}}
\put (550,255){\circle{10}}
\put (550,55){\circle{10}}
\put (441,49){$\times$}
\put (441,249){$\times$}
\put (641,49){$\times$}
\put (641,249){$\times$}

\put (550,155){\vector(1,0){55}}
\put (605,155){\line(1,0){45}}
\put (550,145){\vector(1,0){55}}
\put (605,145){\line(1,0){45}}

\put (550,155){\vector(-1,0){55}}
\put (495,155){\line(-1,0){45}}
\put (550,145){\vector(-1,0){55}}
\put (495,145){\line(-1,0){45}}

\put (545,150){\vector(0,1){55}}
\put (545,205){\line(0,1){50}}
\put (555,150){\vector(0,1){55}}
\put (555,205){\line(0,1){50}}

\put (545,150){\vector(0,-1){55}}
\put (545,95){\line(0,-1){40}}
\put (555,150){\vector(0,-1){55}}
\put (555,95){\line(0,-1){40}}

\put (650,155){\vector(0,1){50}}
\put (650,205){\line(0,1){49}}

\put (650,145){\vector(0,-1){50}}
\put (650,95){\line(0,-1){40}}

\put (450,155){\vector(0,1){50}}
\put (450,205){\line(0,1){49}}

\put (450,145){\vector(0,-1){50}}
\put (450,95){\line(0,-1){40}}

\put (555,255){\vector(1,0){50}}
\put (600,255){\line(1,0){49}}

\put (545,255){\vector(-1,0){50}}
\put (495,255){\line(-1,0){45}}

\put (555,55){\vector(1,0){50}}
\put (600,55){\line(1,0){49}}

\put (545,55){\vector(-1,0){50}}
\put (495,55){\line(-1,0){45}}

\end{picture}

\end{center}

\vspace{-1 cm}
\caption{The RG block transformation in presence of gauge interactions: 
on the left, the paths defining the nearest-neighbors contribution to 
$\Gamma$; on the right, those defining the next-to-nearest-neighbors
contributions.}
\label{fig:blockg}
\vspace{0.5 cm}
\end{figure}

The fermion kernel is also expanded in powers of $A$:
\bea
\Gamma(x_B;\,U(A)) = \Gamma^0(x_B)\:+\:i\,\sum_{x\,r}\,\sum_\mu\,A_\mu(r)\,
\epf_\mu(2x_B-r,x-r)\,\psi_x \nonumber \\
+\sum_{x\,r\,r'}\,\sum_{\mu\nu}\,A_\mu(r)\,A_\nu(r')\,
\eps_{\mu\nu}(2x_B-r,x-r',r-r')\,\psi_x\;. 
\label{gammaexp}
\eea
\noi
The explicit expressions for the functions $\epf_\mu$ and $\eps_{\mu\nu}$ are
displayed in Appendix~\ref{app:b2}.

Inserting the expansion for the coarse action $S'_F$ into the l.h.s. of
Eq.~(\ref{sapof}) 
and that for the fine action $S_F$ and for the fermion kernel ${\cal K}_F$
in its r.h.s.,
with $A^{\pst min}_\mu$ given by (Eq.~(\ref{minfi})),
we obtain a recursion relation for $D'$, $R^{{\pss (1)}\,\prime}_\mu$ and
$R^{{\pss (2)}\,\prime}_{\mu\nu}$. As usual, the recursion for a given 
order depends only on  the solutions for the previous orders. 
Hence, we must solve first the zero order, which is the free fermion 
problem already considered. Inserting the 
solution for the FP propagator $D_{\pss FP}$ into the first order
recursion, we find the FP first order vertex. Then, the FP propagator and
the FP first order vertex determine uniquely (through the second 
order recursion) the FP second order vertex.

\subsection{The first order FP vertex.}
\label{sec:fofpv}

Since in $d=2$ the electric charge has the dimension of a mass,
the lattice coupling constant $g$ defines a relevant direction in the space of
couplings of the interacting theory. At the lowest order, the renormalization
of the coupling constant is trivial:
\be
g'\;=\;2\,g\;\;\;\;\;\;{\textstyle\rm
and}\;\;\;\;\;\;\beta'\;=\;\frac{\beta}{4}\;.
\ee
This can be explicitly seen, since after one RG step 
the first order coarse vertex verifies:
\be
R^{\pss (1)\,\prime}_{\mu}(q=0,q'=0)\;=\;2\,R^{\pss (1)}_{\mu}(q=0,q'=0)\, .
\ee
\noi
To be consistent with the normalization condition (\ref{normal}) we must add
to the RG transformation of Sec.~\ref{sec:fpa} a final step:
\be
A'_\mu (x_B)\;\rightarrow\;\frac{1}{2}\,\,A'_\mu (x_B)\, .\label{renorm}
\ee
\noi
As an effect, the coupling constant is also renormalized.
In the case of asymptotically free theories
$g$ is a marginal coupling, and no additional
renormalization for the gauge field is required when working at tree level.

The recursion relation for the first order vertex is given in
Eq.~(\ref{aprrfo}) of Appendix~\ref{app:c1}.
There, some details about its derivation
are also explained. The explicit form of its solution is a rather 
cumbersome expression, reported in Eq.~(\ref{aprrfos}). Here, we will 
make only a few comments concerning the numerical evaluation of the FP 
first order vertex. 
The iterative solution of the FP equation leads to an expression of the form: 
\be
R^{\pss FP\,(1)}_{\mu}(q,q')\;=\;
\sum_{l,l'=-\infty}^{+\infty}
\,\overline{H}_\mu(q,q',l,l')
\:+\:\sum_{n=1}^\infty\,\frac{1}{4^n}\,\sum_{l,l'=0}^{2^n-1}
\,\overline{X}^{\pss (n)}_\mu(q,q',l,l')\, . \label{fov}
\ee
\noi
We must evaluate numerically the r.h.s. of the last equation for each value of
the momenta $q$ and $q'$.
The first term can be evaluated  with arbitrary (machine)
precision, since the series involved has a very fast convergence.
The second term, however, is more problematic. The function 
$\overline{X}^{\pss (n)}_\mu$ has a complicated structure, 
and the computer time required for the calculation 
of the sums over $l$ and $l'$
grows with the fourth power of $n$. In practice, we must restrict 
the sum over $n$ up to seven terms at most. The effects of this truncation 
are well under control, of order $10^{-5}$.

Problems can arise when truncating the series since gauge invariance
is no more exact. Gauge invariance is guaranteed order by order in perturbation
theory if the vertices satisfy the Ward identities, which for the first order 
vertex read:
\be
\sum_\mu\,\widehat{\left(q+q'\right)}_\mu\,R^{\pss (1)}_\mu(q,q')\;=\;
D^{-1}(q)\:-\:D^{-1}(-q')\, . \label{wifov}
\ee
\noi
This identity is verified by the FP vertex given by Eq.~(\ref{fov}) 
if the sum over $n$ is exactly performed. The truncation of the series 
introduces a violation of the Ward identity, which affects even the continuum
limit. We checked in our calculation that the difference between 
the l.h.s. and the r.h.s. of Eq.~(\ref{wifov}) is of order $10^{-5}$, 
as expected from the convergence properties of the series in $n$. 

To end this subsection, let us study the structure 
of the FP first order vertex.
In two dimensions the lowest dimensional representation of the Dirac algebra
- which we are using in this work - is two-dimensional. The Dirac structure
of the vertex is then highly simplified with respect to four-dimensional
theories. Only four matrices are independent: the identity, $I$, and the Pauli
matrices, related to the Dirac matrices by Eq.~(\ref{diracmat}). Since our RG
transformation violates chiral symmetry, a $\gamma_5$ term is generated in the
vertex through the renormalization steps, besides another term, proportional
to the identity in Dirac space, which was already present in the Wilson
action. The most general FP vertex can be written as
\be
R^{\pss FP\, (1)}_\mu(q,q')\;=\;\sum_\nu\,f^{\pss T}_{\mu\nu} (q,q')\,
\gamma_\nu\:+\:f^{\pss A}_\mu(q,q')\,\gamma_5\:+\:
f^{\pss V}_\mu(q,q')\, I\, , \label{fostr}
\ee
\noi
where the functions $f^{\pss T}_{\mu\nu}$, $f^{\pss A}_\mu$ and 
$f^{\pss V}_\mu$ satisfy the symmetry requirements: hypercubic,
reflection and charge conjugation invariance. 

We found that in the FP vertex all the terms displayed in Eq.~(\ref{fostr}) are
present with almost equal weight.  
In contrast with the four-dimensional case, is here absent the Pauli term
(or ``Clover'' term) involving the interaction with a magnetic field, 
which on the other side cannot exist in $1+1$ dimensions. Here: 
$\sigma_{\mu\nu}=1/2\,[\gamma_\mu,\gamma_\nu]=-i\epsilon_{\mu\nu}\gamma_5$, where
$\epsilon_{\mu\nu}$ is the antisymmetric tensor. The would-be Pauli term is in
fact the $\gamma_5$ term~\cite{wnf}.

\begin{figure}[t]
\centering
\mbox{\psfig{file=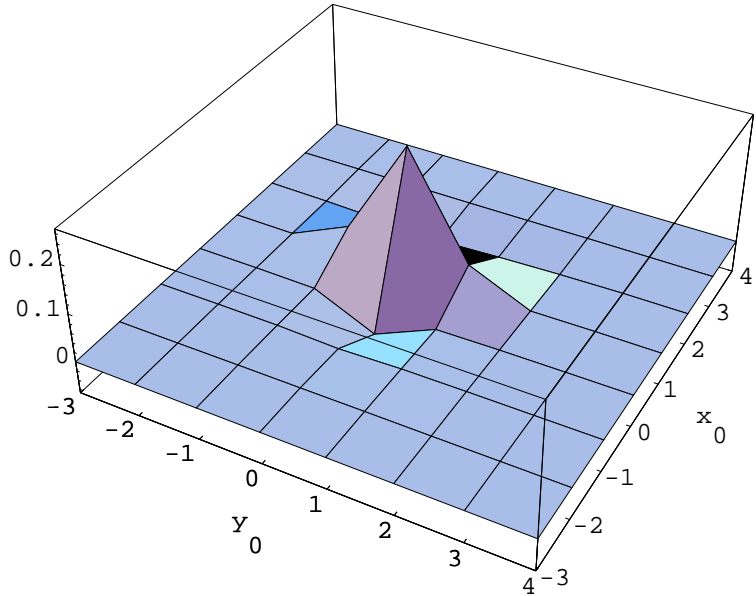,width=10 truecm, angle=0}}
\caption{$f^{\pss T}_{00}(x_0,x_1,0,0)$ on a $8\times 8$ lattice.}
\label{fig:trid}
\vspace{0.5cm}
\end{figure}

\begin{figure}[t]
\centering
\mbox{\psfig{file=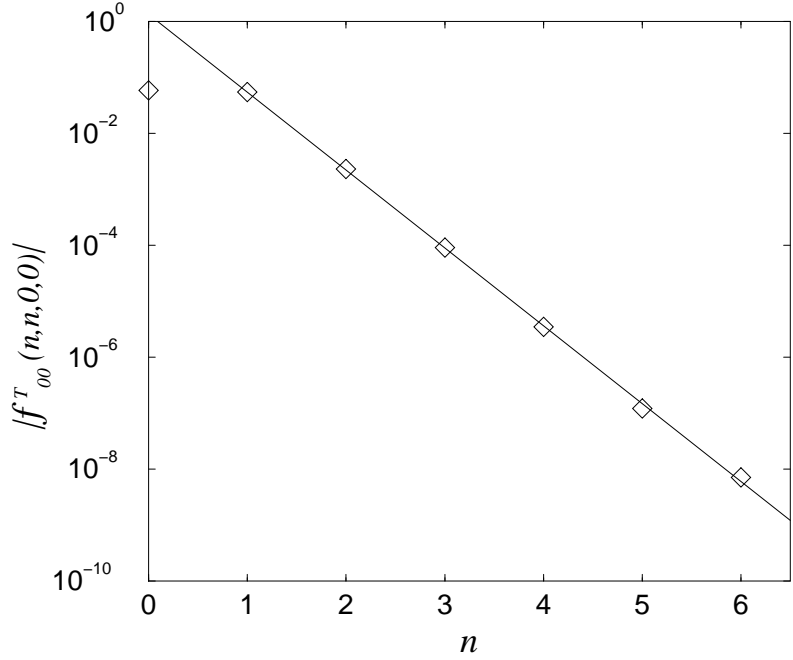,width=10 truecm, angle=-90}}
\vspace{0.2cm}
\caption{Exponential decay of $|f^{\pss T}_{00}(n,n,0,0)|$; 
the continuum line is the best fit $\sim\exp(-n/0.44)$.}
\label{fig:dec}
\vspace{0.5cm}
\end{figure}

The FP vertex has a finite
extension, i.e. it is non-negligible only on a finite set of couplings 
close to the origin. The couplings decay exponentially with the distance, 
with a characteristic length of about $0.5$ lattice units.
In Fig.~\ref{fig:trid} a three-dimensional graph displays the component 
$f^{\pss T}_{00}(x_0,x_1,0,0)$ on a $8\times 8$ lattice, 
while in Fig.~\ref{fig:dec} the exponential decay along the diagonal direction
is reported. The first order vertex gives some indications about the 
interaction range of the full non-perturbative vertex. 
Our results suggest that the couplings confined inside a 
$2\times2$ plaquette are the dominating ones, being those outside this region
smaller than $10^{-3}$ and exponentially decaying.
As a consequence, a good parametrization of the FP action
should be possible with a reasonable number of terms in the lattice 
action\footnote{This is really an important issue 
in view of the four-dimensional theories,
where the Monte Carlo times become non realistic when considering 
lattice actions with complicated coupling structures.}.

\subsection{The second order vertex.}

The fermion propagator and the first order vertex are 
the only ingredients to solve the recursion relation for the second order
vertex. This has a quite complicated expression, as the reader can realize 
by considering the formulas of Appendix~\ref{app:c2}.
It may be worthwhile to display the equation because it is
independent of the particular RG transformation used, the details of which
enter only through the functions $\Theta_{\mu\nu}$,\ $\omega_F$, \ $\epf_\mu$\ 
and $\eps_{\mu\nu}$, whose definition and particular expressions 
for our case are given in the appendices A and B. 
The iterative solution to the FP equation for
the second order vertex is also displayed in Appendix~\ref{app:c2}. It 
can be written in a concise way as 
\bea
&&\hspace*{-1 cm}
R^{\pss (2)\, FP}_{\mu\nu}(q,q',q'')\;=\;-\,i\,
\sum_{l,l',l''=-\infty}^{\infty}\,D_{\pss FP}^{-1}(q)\,
\bar{J}_{\mu\nu}(\tq,\tqp,\tqpp)\,D_{\pss FP}^{-1}(-q') \nonumber \\
&&+\:\sum_{n=1}^{\infty}\,
\left(\frac{1}{2}\right)^{3n-1}\,\sum_{l,l',l''=0}^{2^n-1}\,
D_{\pss FP}^{-1}(q)\,
\bar{I}^{\pss (n)}_{\mu\nu}\,(\tq/2^n,\tqp/2^n,\tqpp/2^n)\,D_{\pss FP}^{-1}(-q')
\nonumber \\
&&-\:R^{\pss (1)\, FP}_\mu(q,-q'')\,D_{\pss FP}(q'')\,
R^{\pss (1)\, FP}_\nu(q'',q')\, . \label{sov}
\eea
\noi
We use the notation $\tq=q+2\pi l$, $\tqp=q'+2\pi l'$ and
$\tqpp=q''+2\pi l''$. The explicit form of the functions $\bar{J}_{\mu\nu}$
and $\bar{I}_{\mu\nu}$ is given in Appendix~\ref{app:c2}.

The matrix structure of the vertex is the most general consistent with the
symmetries. It can be written as
\be
R^{\pss (2)\, FP}_{\mu\nu}(q,q',q'')\;=\;\sum_\rho\,T_{\mu\nu\rho}(q,q',q'')
\,\gamma_\rho\:+\:T_{\mu\nu}^{\pss A}(q,q',q'')\,\gamma_5\:+\:
T_{\mu\nu}^{\pss V}(q,q',q'')\,I\, .
\ee
\noi
As in the case of the first order FP vertex, all the terms in the 
last expression turn out to be of the same order of magnitude.

Again, the sums of the first term in the r.h.s. of Eq.~(\ref{sov}) (the one 
involving $\bar{J}_{\mu\nu}$) can be 
performed with very high precision. The third term gives no difficulty once
the first order FP vertex has been computed.
The second term, however, is an infinite series in $n$, the n-th summand
being a sum of $2^n$ terms which contain very complicated functions,
including the first order vertex on arbitrarily large lattices. Since the 
computation time grows with the sixth power of $n$, we could not
go beyond $n=3$, and even in this case the computation 
of the complete vertex turns out to be 
problematic on a small lattice, due to the large number 
of arguments and the complicated structure of the summands.
However, the perturbative evaluation of the mass gap requires the
second order vertex only for a particular choice of momenta 
and Lorentz indices (see Eq.~(\ref{vp1l})), being as a consequence 
the approximate computation feasible.

The function $\bar{I}^{\pss (n)}$ in Eq.~(\ref{sov}) contains
the first order vertex $R^{\pss (1)}_\mu$ on a lattice of size $2^n
N$, where $N$ is the size of the starting lattice. 
Since we are not able to calculate (at least with good precision) 
the first order vertex on larger lattices, we truncate 
its couplings to relative distances in all components $|\Delta x_\mu|$
smaller than four lattice units.   
This is a quite good approximation, since the couplings associated to
distances outside this region are smaller than $10^{-4}$
(see Fig.~\ref{fig:dec}). We made a numerical check of this approximation:
increasing the truncation size to 6 lattice units the results for 
the second order vertex change by $10^{-5}$ at most. 

Our crudest approximation is the truncation of the series in $n$, this
time at $n=3$. In this case, we expected systematic errors of order $10^{-3}$.
Again, the main danger comes from the violation of gauge invariance, and a
check is given by the Ward identity:
\bea
&&\hspace*{-1 cm}
\sum_\mu\,(\widehat{q-q''})_\mu\,[\,R^{\pss (2)}_{\mu\nu}(q,q'q'')\:+\:
R^{\pss (2)}_{\nu\mu}(q,q'q-q'-q'')\,]\;= \nonumber \\
&&\hspace*{4 cm}
R^{\pss (1)}_\nu(q'',q')\:-\:R^{\pss (1)}_\nu(q,q'+q''-q)\, .
\label{eq:w2}
\eea
\noi
With our approximation the violations of Eq.~(\ref{eq:w2}) 
are at most of order $10^{-3}$. How this systematic error may affect the
computation of the mass gap is unclear at this stage. 

\section{\bf The mass gap.}
\label{sec:mass}

In this Section we review the computation of the 1-loop mass gap 
with both the standard and the FP action.

The 1-loop vacuum polarization is given by
\bea
\Pi_{\mu\nu}(q) &=& \int_{BZ}\,\frac{d^2p}{(2\pi)^2}\,Tr\,\left[\,
R^{\pst (1)}_\mu(p,q-p)\,D(p-q)\,R^{\pst (1)}_\nu(p-q,-p)\,D(p) \right.
\nonumber \\
&+& \left. \left(\,
R_{\mu\nu}^{\pss (2)}(p,-p,p-q)\:+\:R_{\nu\mu}^{\pss (2)}(p,-p,p+q)\,
\right)\,D(p)\,\right]\, . \label{vp1l}
\eea
\noi

\begin{figure}[t]
\centering
\mbox{\psfig{file=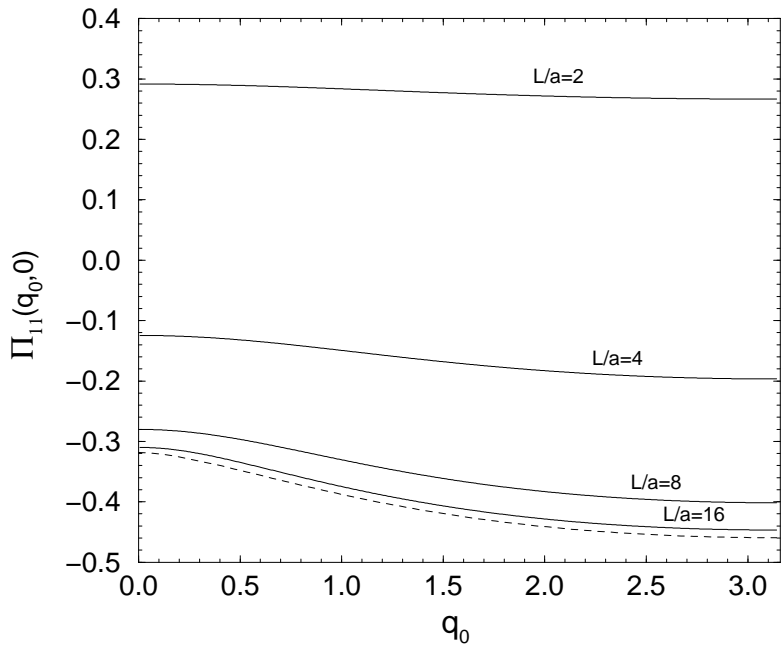,width=10 truecm, angle=-90}}
\caption{The $11$ component of the lattice vacuum polarization $\Pi_{11}$ 
at zero spatial momentum as a function of the
energy for various values of the ratio $L/a$ in the case of the Wilson action;
the extrapolation to infinite volume  (dashed line) is also reported.}
\label{fig:piw}
\vspace{0.5cm}
\end{figure}

\begin{figure}[t]
\centering
\mbox{\psfig{file=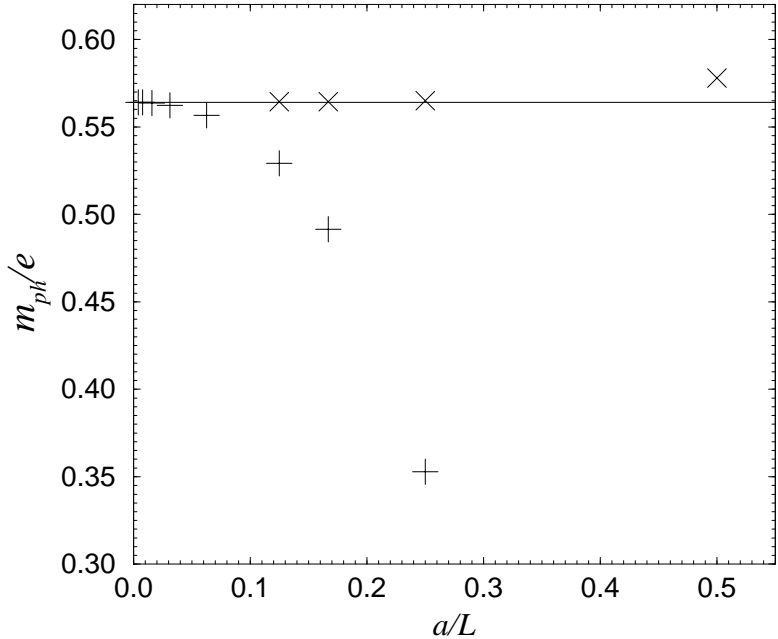,width=10 truecm, angle=-90}}
\caption{The ratio $m_{\pss ph}/e$ as a function of $a/L$ for the
Wilson action (plus) and the FP action (crosses); the solid line is 
the continuum value $1/\sqrt{\pi}$.}
\label{fig:mass}
\vspace{0.5cm}
\end{figure}

Let us now consider the numerical results.
We start discussing those for the standard action.
Fig.~\ref{fig:piw} represents the $11$ component of the lattice 
vacuum polarization $\Pi_{\mu\nu}$ at
zero spatial momentum as a function of the energy $q_0$, 
for different values of the ratio $L/a$; the case $L=\infty$ is also reported.
As expected from the discussion of Sec.~\ref{sec:cuteff}, the cut off effects 
($a$ varying with fixed $L$) show themselves as
finite volume effects ($L$ varying with fixed $a$).
These are large for the Wilson action; in the case $L/a=2$ the theory 
contains no particle at all, since the vacuum polarization has the wrong sign.
The mass is obtained - according to Eq.~(\ref{limit}) - by extrapolating the 
vacuum polarization to zero energy.
In Fig.~\ref{fig:mass} and Table~\ref{tab:mass} we report 
the values of the ratio
$m_{\pss ph}/e$ - in the Table we report also the value of 
$F^{\pss (1)}(a/L)$ (see Eqs.~(\ref{cont}) and (\ref{contino})) ; 
from these data we obtain: 
$m_{\pss ph}/e  = 0.5641900 + 1.9\cdot(a/L)^2 + O((a/L)^4)$, 
to be compared with the
continuum value $m_{\pss ph}^{\pss (c)}/e=0.5641896$.

\begin{table}[t]
\centering
\begin{tabular}{|c|c|c|c|c|}
\cline{1 - 5}
$L/a$ &  $m_{\pss ph}/e\, (Wil)$ & $F^{(1)}\, (Wil)$ & $m_{\pss ph}/e\, (FP)$ &
$F^{(1)}\, (FP)$ \\
\cline{1 - 5}
  2       &            &             &  0.577989475 & 0.013800   \\
  4       &  0.3528    & -0.21132    &  0.565117686 & 0.000928   \\ 
  6       &  0.49158   & -0.07260    &  0.564524513 & 0.000335   \\
  8       &  0.52924   & -0.03495    &  0.564434047 & 0.000244   \\
  16      &  0.55668   & -0.00751    &              &            \\
  32      &  0.56236   & -0.00183    &              &            \\
  64      &  0.56373   & -0.00045    &              &            \\
  128     &  0.56407   & -0.00011    &              &            \\
  256     &  0.56416   & -0.00003    &              &            \\ 
 $\infty$ &  0.5641900 &             &  0.564335184 &            \\
\cline{1 - 5}
\end{tabular}
\caption{the values of $m_{\pss ph}/e$ and $F^{(1)}$ for the Wilson 
and FP action}
\label{tab:mass}
\vspace{0.2cm}
\end{table}

\begin{figure}[t]
\centering
\mbox{\psfig{file=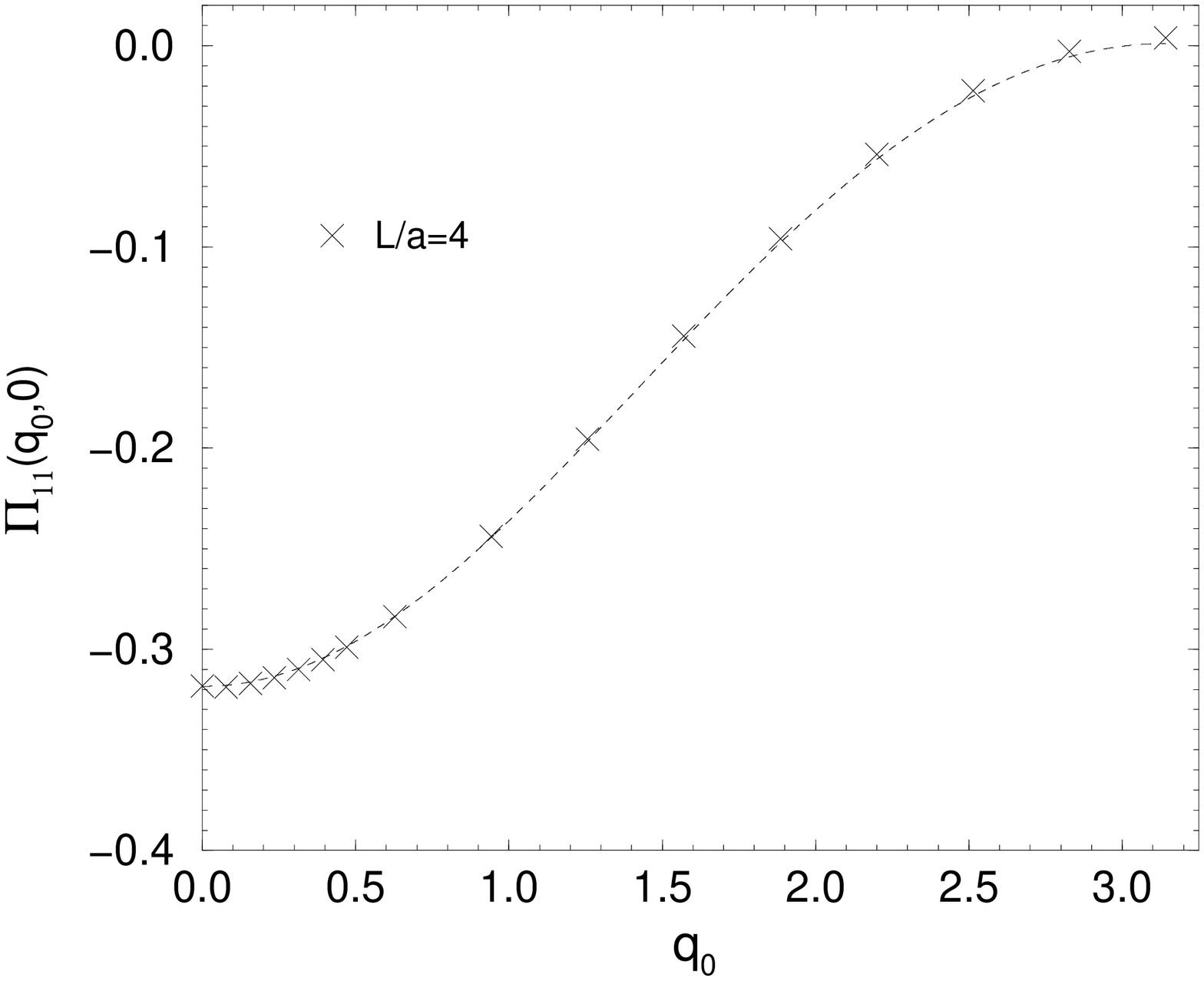,width=10 truecm, angle=-90}}
\caption{The $11$ component of the lattice vacuum polarization 
$\Pi_{11}$ in the infinite volume (dashed line) and for $L/a=4$ (crosses) 
in the case of the FP action.}
\label{fig:pifp}
\vspace{0.5cm}
\end{figure}

In the case of the FP action we observe a radically different behavior.
In Fig.~\ref{fig:pifp} we show the vacuum polarization 
in the infinite volume case,
together with the same quantity in the case $L/a=4$ for some values of the 
energy: we observe only tiny finite volume effects, probably due to 
our numerical approximations.
The cut off effects in the lattice mass gap are directly related 
to the finite volume effects
of the lattice vacuum polarization only for $q_0=0$.
The absence of finite volume effects even for non-zero 
values of the energy with the FP action is in this sense an extra bonus. 
We verified that the volume-independence of the zero spatial 
momentum vacuum polarization 
is indeed a property of the continuum theory, probably related to the fact
that the real underlying theory of the Schwinger model is a free-field 
scalar theory.

In Fig.~\ref{fig:mass} and Table~\ref{tab:mass} we report the results for mass 
gap with the FP action.
Except for the case $L/a=2$, which requires a particular discussion, we see
very small deviations from the continuum, of order of the numerical
errors in the determination of the fixed point vertices 
($10^{-3}\div10^{-4}$). We remark that these deviations cannot 
be considered pure cut off effects, since they are in part 
produced by a violation of
the gauge symmetry. Indeed, a deviation from the correct continuum value 
of order $10^{-4}$ is observed even in the infinite volume case, 
where in principle no cut off effects are present for any action. 
The large deviation from the continuum for $L/a=2$ is related to an additional
effect~\cite{fphn} exponentially decaying with increasing $L/a$ and
related to the finite extension of the FP action.

\section{\bf The 1-loop perfection of the FP action. \label{olpsec}}
\label{sec:perf}

The results of the last Section are consistent with a picture of 
no power-like cut off effects for the 1-loop mass gap 
in the case of the FP action. The point is now whether this outcome
can be interpreted  as an effect of a (hypothetical) 1-loop quantum perfection
of the FP action of the Schwinger model - we recall that any FP
action is by construction perfect only at the {\em classical} level.

The definition of quantum perfection is related to the  behavior 
of the action under RG transformations at finite values of $\beta$. 
We recall the form of the FP action:
\be 
S^{\pss FP}\;=\;\beta\,S_g^{\pss FP}(A)\:+\:
\bar\psi\,\Delta^{\pss FP}(U)\,\psi\, .
\ee
\noi
After one RG transformation the action $S^{\pss FP}$ changes into $S'$:
\be
S'\;=\;\frac{\beta}{4}\,(\,S_g^{\pss FP}(A)\:+\:\delta S_g(A)\,)\:+\:
\bar\psi\,\Delta^{\pss FP}(U)\,\psi
\:+\:\delta S_{\pss F}(\bar\psi,\psi,U)\;.
\label{ass}
\ee
The corrections $\delta S_g$ and  $\delta S_{\pss F}$ have a perturbative
expansion in $g$, the leading term being $O(g^2)$; in particular, 
the fermion correction contains four-fermions interactions.
The 1-loop quantum perfection means the absence of the leading terms 
in the perturbative expansion of $\delta S_g$ and $\delta S_{\pss F}$. 
The absence of cut off
effects in the mass gap for the FP action requires a weaker 
property\footnote{Here we follow the same lines of the argument 
of~\cite{hn1l}.}.
The 1-loop mass gap in units of the charge is a non-universal
function of $a/L$; denoting with a prime the quantities relative 
to the action $S'$ of Eq.~(\ref{ass}), we have:
\be
m^{\pss FP}_{\pss ph}/e\;=\; F^{\pss FP}(a/L)\, ,
\hspace{1 cm}
m'_{\pss ph}/e \;=\; F'(a/L) \;.
\ee
The mass-charge ratio, being the dimensionless ratio of physical quantities,
does not change after a RG transformation. Hence:
\be
F^{\pss FP}(a/L)\;=\;F'(2\,a/L)\;=\;F^{\pss FP}(2\,a/L)\:+\:
\delta F'(2\,a/L)\;. 
\label{difmass}
\ee
\noi
The last equality is suggested by Eq.~(\ref{ass}).
The correction $\delta F'$ comes from the leading terms in the perturbative 
expansion of $\delta S_g$ and  $\delta S_{\pss F}$; absence of cut off
effects for the FP action means  $\delta F'=0$, 
since we see from Eq.~(\ref{difmass}) 
that $F^{\pss FP}$ is in this case independent from $a/L$.
After inspecting all the possible vertices generated by  the RG transformation,
one concludes that the only term which can contribute at 1-loop 
is a tree level term, quadratic\footnote{One can see that, for example,
vertices containing four fermion - or higher dimensional - interactions 
cannot contribute.} in the field $A_\mu$. 
Gauge invariance implies (in two dimensions) the form:
\be
\sum_{\mu\nu}\,A_\mu(q)^*\,C(q)\,\left(\,\mq^2\,\delta_{\mu\nu}\:-\:
\hq_\mu\,\hq_\nu^*\,\right)\,A_\nu(q)\, ,
\ee
\noi
where $C(q)$ is some regular function\footnote{The regularity 
of $C(q)$ follows from the general theory of the RG.} of $q$.
This term gives no contribution to the mass gap, since it vanishes at $q=0$.
As a consequence, $\delta F'\,=\,0$.

The previous discussion shows that the absence of cut off effects in the 
1-loop mass gap is indeed an effect of the {\em classical} perfection
of the FP action\footnote{In the framework of the $O(3)-\sigma$ model,
this conclusion is also suggested \cite{pisa} \cite{pisa1}
by the observation that the tree-level on-shell Symanzik improvement
is fixed by the cancellation of the $O(a^2)$ cut off
effects in the 1-loop contribution to the mass gap.}. 
This is a non trivial point, since a formal 
argument~\cite{wil,hnymp} of the RG implies the 1-loop quantum perfection 
of the FP action. This argument has been recently disproved by 
Hasenfratz and Niedermayer~\cite{hn1l} by the explicit determination in
perturbation theory of 
the 1-loop quantum perfect action of the 
$O(3)-\sigma$ model. We do not expect something different for
the Schwinger model, although, due to the different nature of the lagrangian
(gauge interactions between two fields instead of self-interactions) we think
it would be worthwhile to repeat the check even in the present case.
Another important point would be to understand {\em where} the formal
argument of the RG breaks down.
\section{\bf Conclusions.}
\label{sec:conc}
Let us briefly summarize the results and the conclusions of this work. 
We have obtained perturbatively the FP action of the Schwinger model
for a particular RG transformation. Using the non-compact 
formulation, we could solve analytically the pure gauge part. 
In this way we could concentrate the numerical effort 
to the fermion sector, which was treated perturbatively. 

The photon-fermion first order vertex turns out to have couplings exponentially
decaying with the distance, as expected from RG arguments.
The most important ones are those connecting first and second neighbors.
We expect from this observation that a good approximation 
of the FP action can be obtained
with a simple parametrization containing only short-ranged couplings.

The perfection of the FP action has been shown by computing the
mass gap at finite spatial volume. We found big cut off effects with the
standard  action, and negligible with the FP action. Contrarily to what can be
naively inferred, the calculation of the mass gap represents
a check of the classical - tree level - 
perfection, and not of the 1-loop quantum perfection; 
this latter property can be checked~\cite{hn1l} 
by computing higher excited states in the spectrum.

\vspace{1cm}

{\bf Acknowledgments} 

We are indebted with P.~Hasenfratz and F.~Niedermayer for having 
introduced us in the subject and for useful suggestions. 
We acknowledge valuable discussions
with C.B.~Lang and T.K.~Pany. The work received financial 
support from Fondazione ``A.~Della~Riccia''- Italy (F.F.) and Ministerio 
de Educaci\'on y Cultura - Spain under grant PF-95-73193582 (V.L.).

\appendix

\section{\bf Appendix.}
\label{app:a}

In this Appendix we give some details about the solution of the recursion
relation for the free gauge field.
We follow the ideas of~\cite{hnymp}, and we refer to this paper
for further details.

\subsection{The FP pure gauge propagator.}
\label{app:a1}

In this part we give some hints about the algebraic manipulations which
lead to the pure gauge FP propagator, Eq.~(\ref{fpgapro}).
We write the general formulas for an arbitrary
dimension $d$, specializing at the end the solutions to $d=2$.

Consider the RG kernel:
\be
{\cal K}_g(A',A)\;=\;\sum_{x_B\,\mu}\,\left(\,A'_\mu(x_B)\,-\,\sum_{x\,\nu}\,
\omega_{\mu\nu}(2x_B-x)\,A_\nu(x)\,\right)^2\, .
\ee
\noi
In the case of the transformation of Eq.~(\ref{gaugekernel}), 
generalized to arbitrary dimensions, $\omega_{\mu\nu}$ is written:
\be
\omega_{\mu\nu}(2
x_B-x)\;=\;2^{d_A-1}\,\delta_{\mu\nu}\,\left(\,\delta_{2x_B\, ,\, x}\:+\:
\delta_{2x_B+\hat{\mu}\, ,\, x}\,\right)\, ,
\ee
\noi
where $d_A=(d-2)/2$ is the dimension of the gauge field.
In Fourier space the last formula reads:
\be
\omega_{\mu\nu}(q)\;=\;\sum_x\,e^{-iqx}\,\omega_{\mu\nu}(x)\;=\;2^{d_A-1}\,
\left(\,1\,+\,e^{iq_\mu}\,\right)\,\delta_{\mu\nu}\, .
\label{eq:omegag}
\ee

Since the original action and the RG kernel are quadratic in the gauge field,
the coarse action is also quadratic. From gaussian integration we obtain a
recursion relation which relates the coarse and fine propagators~\cite{hnymp}:
\be
G'_{\mu\nu}(q)\;=\;2^{-d}\,\sum_{l=0}^1\,\sum_{\rho\sigma}\,
\omega_{\mu\rho}(\frac{q+2\pi l}{2})\,
G_{\rho\sigma}(\frac{q+2\pi l}{2})\,
\omega^\dagger_{\sigma\nu}(\frac{q+2\pi l}{2})\:+\:
\frac{1}{\kappa_g}\,\delta_{\mu\nu}\, .
\ee
\noi
After $n$ iterations of the recursion relation one gets
\bea
G^{\pss (n)}_{\mu\nu}(q) &=& 2^{(2d_A-d)n}\,
\sum_{l=0}^{2^n-1}\,
\sum_{\rho\sigma}\,\Omega^{\pss (n)}_{\mu\rho}(\frac{q+2\pi l}{2^n})\,
G^{\pss (0)}_{\rho\sigma}(\frac{q+2\pi l}{2^n})\,
\Omega^{{\pss (n)}\,\dagger}_{\sigma\nu}(\frac{q+2\pi l}{2^n}) \nonumber \\
&+& \frac{1}{\kappa_g}\,\sum_{j=0}^{n-1}\,2^{(2d_A-d)j}\,
\sum_{l=0}^{2^j-1}\,\sum_\rho\,
\Omega^{\pss (j)}_{\mu\rho}(\frac{q+2\pi l}{2^j})\
\Omega^{{\pss (j)}\,\dagger}_{\rho\nu}(\frac{q+2\pi l}{2^j})\, ,
\label{gaugeiter}
\eea
\noi
where
\be
\Omega^{\pss (j)}_{\mu\nu}(\frac{q+2\pi l}{2^j})\;=\;2^{-d_Aj}\,\left[\,
\omega(\frac{q+2\pi l}{2})\,\omega(\frac{q+2\pi l}{2^2})\ldots
\omega(\frac{q+2\pi l}{2^j})\,\right]_{\mu\nu}
\ee
\noi
and
\be
\Omega^{\pss (0)}_{\mu\nu}(q+2\pi l)\;=\;\delta_{\mu\nu}\, .
\ee
In our case (Eq.~(\ref{eq:omegag})) the result is:
\be
\Omega^{\pss (j)}_{\mu\nu}(\frac{q+2\pi l}{2^j})\;=\;
2^{-j}\,\frac{e^{iq_\mu}-1}
{\exp(i\frac{q_\mu+2\pi l_{\mu}}{2^j})-1}\,\delta_{\mu\nu}\ .
\ee
\noi

The FP propagator is obtained by inserting this last expression into
Eq.~(\ref{gaugeiter}) and taking the $n\rightarrow\infty$ limit. 
In the r.h.s. of Eq.~(\ref{gaugeiter}), only the modes with 
$\frac{q+2\pi l}{2^n}\sim\frac{1}{2^n}$ contribute to the homogeneous 
part (the one containing $G_{\mu\nu}^{(0)}$);
the inhomogeneous term (proportional to $\kappa_g^{-1}$) 
is easily calculated by making use of
following formula, valid for one-dimensional summation:
\be
\sum_{j=0}^{2^n-1}\,\frac{1}{4\,\sin^2(\frac{q+2\pi j}{2^{n+1}})}\;=\;
\frac{2^{2n}}{4\sin^2(q/2)} \, .
\ee
\noi 
The result for the FP propagator is, in two 
dimensions\footnote{As may be directed checked
using the above displayed formulas, our RG transformation converges 
to a FP only for $d<3$.}:
\be
G_{\mu\nu}^{\pss FP}(q)\;=\;\left(\,f_\mu(q)\:+\:\frac{2}{\kappa_g}\,\right)\,
\delta_{\mu\nu}\;+\;(\xi^{-1}\,-\,1)\,g(q)\,\hq_\mu\,\hq_\nu^* \, ,
\ee
\noi
where
\bea
f_\mu(q) &=& \left|\,\hq_\mu\,\right|^2\,\sum_{l=-\infty}^{+\infty}\,
\frac{1}{\left(q_\mu+2\pi\, l_\mu\right)^2}\,
\frac{1}{\left(q+2\pi\,l\right)^2} \nonumber \\
g(q) &=& \sum_{l=-\infty}^{+\infty}\,\frac{1}{\left(q+2\pi\,l\right)^4}\, .
\label{apfgeq}
\eea
\noi

\subsection{The connecting tensor.}
\label{app:a2}

According to the discussion of Section~\ref{sec:fpa}, 
in order to solve the problem of
the interaction between gauge field and fermions, we need the fine
configuration minimizing the r.h.s. of Eq.~(\ref{sapo})
as a function of the coarse configuration $A^\prime$. 
Since both the fine action and the gauge kernel are
quadratic, the relation is linear:
\be
A_\mu^{min}(\frac{q+2\pi\,l}{2})\;=\;\sum_\nu\,Z_{\mu\nu}(\frac{q+2\pi\,l}{2})
\,A_\nu^\prime(q)\, ,
\ee
\noi
where $l_\mu=0,1$.  In ref.~\cite{hnymp} is shown that
\be
Z_{\mu\nu}(\frac{q+2\pi l}{2})\;=\;\sum_{\rho\sigma}\,
G^{\pss FP}_{\mu\rho}(\frac{q+2\pi l}{2})\,
\omega_{\rho\sigma}^\dagger(\frac{q+2\pi l}{2})\,
G^{{\pss FP}\,-1}_{\sigma\nu}(q)\ .
\ee

In the iterative solution of the recursion relations for the vertices, 
some products of $Z_{\mu\nu}$ appear. Since these products have always the
same form, it is useful to define
\be
\Theta^{\pss (n)}_{\mu\nu}(q/2^n)\;=\;2^{-(d_A+2)n}\,
\sum_{\rho_1\ldots\rho_{n-1}}\,Z_{\mu\rho_1}(\frac{q}{2^n})\,
Z_{\rho_1\rho_2}(\frac{q}{2^{n-1}})\,\cdots\,Z_{\rho_{n-1}\nu}(\frac{q}{2})
\, .
\ee
\noi

Particularizing to our transformation and fixing $d=2$,  
$Z_{\mu\nu}$ and $\Theta_{\mu\nu}^{\pss (n)}$ read:
\bea
Z_{00}(q) &=& \f{1}{2}\,\left(\,1\,+\,e^{-iq_1}\,\right)\,
\left[\,\left|\hq_0\right|^2\,\frac{g(q)}{g(2q)}\,f_1(2q)\:+\:
\left|\widehat{2q_1}\right|^2\,f_0(q)\,\right]
\nonumber \\
Z_{01}(q) &=& \f{1}{2}\,\hq_0\,\widehat{2q}_2^*\,
\left[\,\frac{g(q)}{g(2q)}\,f_0(2q)\:-\:
f_0(q)\,\frac{\left|\widehat{2q}_0\right|^2}{\left|\hq_0\right|^2}\,
\right]\, , 
\eea
\noi
and
\bea
\Theta^{\pss (n)}_{00}(q/2^n) &=&\frac{1}{8^n}\,
\frac{\hq_0^*}{\widehat{(q_0/2^n)}^*}\,\left[\,|\hq_1|^2\,f_0(q/2^n)\:+\:
\left|\widehat{(q_0/2^n)}\right|^2\,\frac{g(q/2^n)}{g(q)}\,f_1(q)\,\right]
\nonumber \\
\Theta^{\pss (n)}_{01}(q/2^n) &=&\frac{1}{8^n}\, 
\frac{\hq_1^*}{\widehat{(q_0/2^n)}^*}\,\left[\,\left|\widehat{(q_0/2^n)}\right|^2
\,\frac{g(q/2^n)}{g(q)}\,f_0(q)\:-\:|\hq_0|^2f_0(q/2^n)\,\right]\, .
\nonumber\\
\;\;\;\;\;
\eea
\noi
The remaining components can be obtained from these by properly changing
$0\leftrightarrow 1$.

When $n\rightarrow\infty$, keeping $q$ fixed, we obtain:
\bea
\Theta^{\pss (\infty)}_{00}(q) &=& i\,\frac{\hq_0^*}{q_0}\,\left(\,
\frac{|\hq_1|^2}{q^2}\:+\:\frac{q_0^2}{q^4}\,\frac{f_1(q)}{g(q)}\,\right)
\nonumber \\
\Theta^{\pss (\infty)}_{01}(q) &=& i\,\frac{\hq_1^*}{q_0}\,\left(\,
\frac{q_0^2}{q^4}\,\frac{f_0(q)}{g(q)}\:-\:\frac{|\hq_0|^2}{q^2}\,\right)\, .
\eea
These expressions appear in the solution of the recursion relations for the
first and second order vertices, as given in Appendix~\ref{app:c}.

\section{\bf Appendix.}
\label{app:b}

In this Appendix we give some details concerning the fermion RG
transformation. We start with some relations concerning the free fermion
problem wich are relevant for the computation of the FP vertices.
In a second part we work out the fermion kernel in presence of the
interaction with a $U(1)$ gauge field.

\subsection{Formulas for the free fermion problem.}
\label{app:b1}

The free fermion kernel is defined through a function $\omega_F$
\be
\Gamma^0(x_B)\;=\;\sum_x\,\omega_F(2x_B-x)\,\psi_x\, .
\ee
\noi
The RG transformation of Sec.~\ref{subsec:freef} reads 
in arbitrary dimensions:
\bea
\Gamma^0(x_B)&=&2^{d_{\psi}-d}\,\left[\,\psi(2x_B)\:+\:
\frac{1}{2}\,\sum_\mu\:\sum_{\lambda=\pm 1}\,\psi(2x_B\,+\,\lambda\hat\mu)
\right. \nonumber \\
&+&\frac{1}{4}\,\sum_{\mu<\nu}\:\sum_{\lambda_\mu=\pm 1}\:\sum_{\lambda_\nu=\pm 1}\,
\psi(2x_B\,+\,\lambda_\mu\hat{\mu}\,+\,\lambda_\nu\hat{\nu}) \nonumber\\
&+&\left.\cdots+ \frac{1}{2^d}\,\sum_{\lambda_1=\pm 1}\cdots\sum_{\lambda_d=\pm 1}\,
\psi(2x_B\,+\,\lambda_1\hat{1}\,+\cdots\,\lambda_d\hat{d})\,\right]\, ,
\label{apfreefk}
\eea
\noi
where $d_{\psi}=(d-1)/2$ is the dimension of the fermion field.
In Fourier transform:
\be
\omega_F(q)\;=\;2^{d_\psi}\prod_{\mu}\,\cos^2\frac{q_\mu}{2}\, .
\ee

For the iterative solution of the FP equations, it  
is useful to define a new function, analogous to $\Omega^{\pss (n)}_{\mu\nu}$
of the pure gauge problem:
\be
\Omega^{\pss (n)}_F(q/2^n)\;=\;2^{-d_\psi\,n}\,
\omega_F(q/2)\,\omega_F(q/2^2)\,\cdots\,\omega_F(q/2^n)\, .
\ee
\noi
In the case of the transformation~(\ref{apfreefk}):
\be
\Omega^{\pss (n)}_F(q/2^n)\;=\;\prod_\mu 2^{-2n}\,
\frac{\sin^2 (q_\mu/2)}{\sin^2 (q_\mu/2^{n+1})}\, .
\ee
Using ({\em mutatis mutandis}) Eq.~(\ref{gaugeiter}) one arrives,
in two dimensions\footnote{It can be easily checked that this RG transformation
converges to a FP for $d\leq4$.}, to the result of
Eqs.~(\ref{alphafp}) and (\ref{betafp}).

\subsection{The fermion gauge invariant kernel.}
\label{app:b2}

In this part we shall write down some formulas concerning the fermion
kernel in presence of gauge interactions. 
In order to keep the coarse action gauge invariant, we define a gauge 
covariant average procedure for the fermion fields defined on the original 
fine lattice. This is achieved by parallel transporting 
the fine fields to the coarse site to which
they contribute. We choose the simplest symmetric paths which join $x_B$ with
its neighbors to make (\ref{apfreefk}) gauge covariant (we
write the formulas directly for the case $d=2$):
\be
\Gamma(x_B;\,U)\;=\;\frac{\sqrt{2}}{4}\,\left\{\,\psi(2x_B)\:+\:\frac{1}{2}\,
\Gamma_{fn}(x_B;\,U)\:+\:\frac{1}{4}\,\Gamma_{dn}(x_B;\, U)\,\right\}\, ,
\label{apfk}
\ee
\noi
where $\Gamma_{fn}$ and $\Gamma_{dn}$ stand for the contribution of the first
and diagonal neighbors respectively, and are built from the paths depicted in
Fig.~\ref{fig:blockg}. The expression for them are:
\bea
\Gamma_{fn}(x_B;\, U)&=&U_0(2x_B)\psi(2x_B+\hat{0})\,+\,
U_0^\dagger(2x_B-\hat{0})\,\psi(2x_B-\hat{0}) \nonumber \\ 
&+& U_1(2x_B)\psi(2x_B+\hat{1})\,+\,U_1^\dagger(2x_B-\hat{1})\,
\psi(2x_B-\hat{1}) \,.\nonumber \\
\eea
\noi
and 
\bea
& &\Gamma_{dn}(x_B;\, U)\;= \nonumber \\
& &\left[\,U_0(2x_B)U_1(2x_B+\hat{0})\,+\,
U_1(2x_B)U_0(2x_B+\hat{1})\,\right]\,\psi(2x_B+\hat{0}+\hat{1})\:+ 
\nonumber \\
&+&\left[\,U_0(2x_B)U_1^\dagger(2x_B+\hat{0}-\hat{1})\,+\,
U_1^\dagger(2x_B-\hat{1})U_0(2x_B-\hat{1})\,\right]\,
\psi(2x_B+\hat{0}-\hat{1}) \nonumber \\
&+&\left[\,U_0^\dagger(2x_B-\hat{0})U_1(2x_B-\hat{0})\,+\,
U_1(2x_B)U_0^\dagger(2x_B-\hat{0}+\hat{1})\,\right]\,
\psi(2x_B-\hat{0}+\hat{1})\ \nonumber \\
&+&\left[ U_0^\dagger(2x_B-\hat{0})U_1^\dagger(2x_B-\hat{0}-\hat{1})
+U_1^\dagger(2x_B-\hat{1})U_0^\dagger(2x_B-\hat{0}-\hat{1})\right]
\times \nonumber \\
& &\;\;\;\;\;\;\;\;\;\;\;\;\;\;\;\;\;\;\;\;\;\;\;\;\;\;\;\;\;\;\;\;\;\;\;
\;\;\;\;\;\;\;\;\;\;\;\;\;\;\;\;\;\;\;\;\;\;\;\;\;\;\;\;\;\;\;\;\;\;\;\;\;
\psi(2x_B-\hat{0}-\hat{1})\, .
\eea
\noi
Expanding $U_\mu$ in powers of $A_\mu$ we obtain:
\bea
\hspace*{-0.75 cm}
\Gamma(x_B;\,U(A)) &=& \Gamma^0(x_B)\:+\:i\,\sum_{x\,r}\,\sum_\mu\,A_\mu(r)\,
\epf_\mu(2x_B-r,x-r)\,\psi_x \nonumber \\
&+&\sum_{x\,r\,r'}\,\sum_{\mu\nu}\,A_\mu(r)\,A_\nu(r')\,
\eps_{\mu\nu}(2x_B-r,x-r',r-r')\,\psi_x\,+\,O(A^3)\,.
\nonumber \\
\label{apgammaexp}
\eea
\noi
In momentum space $\epf_\mu$ and $\eps_{\mu\nu}$ read:
\be
\epf_0(q,q')\;=\;\frac{\sqrt{2}}{8}\,\left(\,\hat{q'}_0^*\:-\:\hq_0^*\,\right)
\,\left(1\:+\:\frac{\cos q_1\,+\,\cos q'_1}{2}\,\right)
\ee
\noi
and
\bea
\hspace*{-1.5 cm}
\eps_{00}(q,q',q'') &=& -\,\frac{\sqrt{2}}{16}\,
\left(\,e^{-iq_0}\:+\:e^{-iq_0'}\,\right)
\,\left(1\:+\:\frac{\cos q_1\,+\,\cos q_1'}{2}\,\right)
\nonumber \\
\eps_{01}(q,q',q'') &=& \frac{\sqrt{2}}{16}\,
\exp(-i\frac{q_0-q_0''}{2})
\,\exp(-i\frac{q_1'+q_1''}{2}) \times \nonumber
\eea
\be
\hspace*{0.5 cm}
\left(\sin\frac{q_0+q_0''}{2}\sin\frac{q_1''-q_1'}{2}
\,-\,\sin(q_0'-\frac{q_0-q_0''}{2})\sin(q_1-\frac{q_1'+q_1''}{2})\right)\, .
\label{epstensor}
\ee
\noi
The remaining components can be obtained as usual by properly changing
$0\leftrightarrow 1$.

\section{\bf Appendix.}
\label{app:c}

In this Appendix we write down the equations relative to the RG iteration
relation for the first and second order vertex. They are the result of
a long but straightforward calculation.

\subsection{Recursion relation for the first order vertex.}
\label{app:c1}

In order to simplify the expressions, we use the notation 
$\tilde{q}=q+2\pi l$ and $\tilde{q'}=q'+2\pi l'$.
The recursion relation verified by the first order vertex is:
\bea
R_\mu^{{\pss (1)}\, \prime}(q,-q') &=&
D_{\pss FP}^{-1}(q)\,\left(\frac{1}{2}\right)^{2d}\,
\sum_{l,l'=0}^1\,\sum_\nu\,
Z_{\nu\mu}(\frac{\tilde{q}-\tilde{q}'}{2})\,\times \nonumber \\
&& \left[\,H_\nu(\tq/2,\,\tqp/2)\:+\:
X_\nu(\tq/2,\,\tqp/2)\,\right]\,D_{\pss FP}^{-1}(q')\, ,
\label{aprrfo}
\eea
\noi 
where
\be
H_\nu(k,k')\;=\;\omega_F(k)\,D_{\pss FP}(k)\,R^{\pss (1)}_\nu(k ,\, -k')\,
D_{\pss FP}(k')\,\omega_F^\dagger(k')
\ee
\noi
and
\be
X_\nu(k,k')\;=\;\omega_F(k)\,D_{\pss FP}(k)\,
{\cal E}^{{\pss (1)}\,\dagger}_\nu(k',\,-k)\:-\:\epf_\nu(k,\,-k')\,
D_{\pss FP}(k')\,\omega_F^\dagger(k')\, .
\ee
\noi
We point out that this formula holds formally for arbitrary dimensions
and gauge groups\footnote{The RG does not act on the color indices.}.

We now specialize derivation to the case of the Schwinger 
model\footnote{The derivation
for the general case, including also 4d non-abelian gauge theories, 
goes essentially on the same lines.}. In this case a factor-two 
renormalization of the gauge field is required (see the discussion
of Sec.~\ref{sec:fofpv}). The recursion relation (\ref{aprrfo})
leads to the FP vertex:
\be
R^{\pss FP\,(1)}_{\mu}(q,-q')\;=\;
\sum_{l,l'=-\infty}^{+\infty}\,\overline{H}_\mu(q,q',l,l')\:+\:
\sum_{n=1}^\infty\,\frac{1}{4^n}\,\sum_{l,l'=0}^{2^n-1}
\,\overline{X}^{\pss (n)}_\mu(q,q',l,l')\, ,
\label{aprrfos}
\ee
\noi
where
\bea
&&\hspace*{-1cm}
\overline{H}_\mu(q,q',l,l')\;=\;-\,D_{\pss FP}^{-1}(q)\,\sum_{\rho\beta\omega}\,
\Theta^{\pss (\infty)}_{\rho\mu}(\tilde{q}-\tilde{q}')\,
\frac{\tilde{q}_\beta\,\tilde{q}'_\omega}{\tilde{q}^2\,\tilde{q}^{\prime\,2}}\,
\gamma_\beta\,\gamma_\rho\,\gamma_\omega\,D_{\pss FP}^{-1}(q')
\,\times \nonumber \\
&&\hspace*{4 cm}
\prod_{\alpha}\,\frac{16\sin^2(q_\alpha/2)\sin^2(q'_\alpha/2)}
{\tilde{q}_\alpha^2\,\tilde{q}_\alpha^{\prime\,2}}
\eea
\noi 
and
\bea
&&\hspace*{-2 cm}
\overline{X}^{\pss(n)}_\mu(q,q',l,l')\;=\;
\frac{1}{\sqrt{2}}\,D_{\pss FP}^{-1}(q)\,\sum_\rho\,
\Theta_{\rho\mu}^{\pss (n)}(\frac{\tilde{q}-\tilde{q}'}{2^n})\times \nonumber \\
&& \hspace*{-2 cm}
\left[\,\Omega_F^{\pss (n)}(\tilde{q}/2^n)\,
D_{\pss FP}(\tilde{q}/2^n)\,
{\cal E}^{{\pss (1)}\,\dagger}_\rho(\tilde{q}'/2^n\, ,\,
-\tilde{q}/2^n)\,
\Omega_F^{{\pss (n-1)}\, \dagger}(\tilde{q'}/2^{n-1})\,
\:-\right. \nonumber \\
&& \hspace*{-2 cm}
\left.
\Omega_F^{\pss (n-1)}(\tilde{q}/2^{n-1})\,
 \epf_\rho(\tilde{q}/2^n\, ,\,-\tilde{q}'/2^n)\,
D_{\pss FP}(\tilde{q}'/2^n)\,
\Omega_F^{{\pss (n)}\, \dagger}(\tilde{q}'/2^n)\,\right]\,
D_{\pss FP}^{-1}(q') \, . 
\eea

\subsection{Recursion relation for the second order vertex.}
\label{app:c2}

The recursion relation for the second order vertex is much more complicated,
and it involves the FP first order vertex:
\bea
&&R_{\mu\nu}^{{\pss (2)}\,\prime}(q,q',q'')\;=\;\frac{1}{2^{3d}}\,
\sum_{l,l',l''=0}^1\,\sum_{\rho\sigma}\,
\left\{\,D_{\pss FP}^{-1}(q)\right. \,Z_{\rho\mu}(\frac{\tq-\tqpp}{2})\,
Z_{\sigma\nu}(\frac{\tqp+\tqpp}{2})\,\times \nonumber \\
&&\hspace*{1 cm}\left. \left[\,J_{\rho\sigma}(\tqm ,\tqpm ,\tqppm)
\;+\;I_{\rho\sigma}(\tqm ,\tqpm ,\tqppm)\,\right]\,
D_{\pss FP}^{-1}(-q')\,\right\}\nonumber \\
&& \hspace*{1 cm}
-\:R^{\pss (1)\, FP}_\mu(q,-q'')\,D_{\pss FP}(q'')\,
R^{\pss (1)\, FP}_\nu(q'',q')\;,
\eea
\noi
where the functions $H_{\rho\sigma}$ and $I_{\rho\sigma}$ are:
\be
J_{\rho\sigma}(k,k',k'')\;=\;\omega_F(k)\,D_{\pss FP}(k)\,
R_{\rho\sigma}^{\pss (2)}(k,k',k'')\,D_{\pss FP}(-k')\,\omega_F^\dagger(-k')
\ee
\noi
and
\bea
&&\hspace{-1 cm} I_{\rho\sigma}(k,k',k'')\;=\nonumber \\
&&-\:\eps_{\rho\sigma}(k,k',k'')\,D_{\pss FP}(-k')\,\omega_F^\dagger(-k') 
\:-\:\omega_F(k)\,D_{\pss FP}(k)\,
{\cal E}^{{\pss (2)}\,\dagger}_{\rho\sigma}(-k,-k',-k'') \nonumber \\
&&+\:\omega_F(k)\,D_{\pss FP}(k)\,R^{\pss (1)\, FP}_\rho(k,-k'')\,
D_{\pss FP}(k'')\,R^{\pss (1)\, FP}_\sigma(k'',k')\,
D_{\pss FP}(-k')\,\omega_F^\dagger(-k') \nonumber \\
&&+\:\omega_F(k)\,D_{\pss FP}(k)\,R^{\pss (1)\, FP}_\rho(k,-k'')\,
D_{\pss FP}(k'')\,{\cal E}^{{\pss (1)}\,\dagger}_\sigma(-k',-k'')
\nonumber \\
&&-\:\epf_\rho(k,-k'')\,D_{\pss FP}(k'')\,R^{\pss (1)\, FP}_\sigma(k'',k')
\,D_{\pss FP}(-k')\,\omega_F^\dagger(-k') \nonumber \\
&&-\:\epf_\rho(k,-k'')\,D_{\pss FP}(k'')\,
{\cal E}^{{\pss (1)}\,\dagger}_\sigma(-k',-k'')\, .
\eea

The FP of the above displayed recursion relation is (taking again into
account the factor-two renormalization of the gauge field): 
\bea
&& \hspace*{-1 cm}
R^{\pss (2)\, FP}_{\mu\nu}(q,q',q'')\;=\;-\,i\,
\sum_{l,l',l''=-\infty}^{+\infty}\,D_{\pss FP}^{-1}(q)\,
\bar{J}_{\mu\nu}(\tq,\tqp,\tqpp)\,D_{\pss FP}^{-1}(-q') \nonumber \\
&&+\:\sum_{n=1}^{+\infty}\,
\left(\frac{1}{2}\right)^{3n-1}\,\sum_{l,l',l''=0}^{2^n-1}\,
D_{\pss FP}^{-1}(q)\,
\bar{I}^{\pss (n)}_{\mu\nu}\,(\tq/2^n,\tqp/2^n,\tqpp/2^n)\,
D_{\pss FP}^{-1}(-q')
\nonumber \\
&&-\:R^{\pss (1)\, FP}_\mu(q,-q'')\,D_{\pss FP}(q'')\,
R^{\pss (1)\, FP}_\nu(q'',q')\, ,
\eea
\noi
where
\bea
&&\hspace{-1cm}
\bar{J}_{\mu\nu}(k,k',k'')\;=\;\sum_{\rho\sigma}\,
\Theta^{\pss (\infty)}_{\rho\mu}(k-k'')\,
\Theta^{\pss (\infty)}_{\sigma\nu}(k'+k'')\,
\sum_{\beta\,\delta\,\omega}\,\frac{k_\beta}{k^2}\,
\frac{k'_\delta}{k^{\prime\prime\, 2}}\,
\frac{k'_\omega}{k^{\prime\, 2}}\,
\gamma_\beta\,\gamma_\rho\,\gamma_\delta\,\gamma_\sigma\,\gamma_\omega
\,\times \nonumber \\
&&\hspace{2 cm}
\prod_\alpha\,\frac{16\sin^2(k_\alpha/2)\sin^2(k'_\alpha/2)}
{k_\alpha^2\;k_\alpha^{\prime\, 2}}
\eea
\noi
and
\bea
&&\hspace{-1 cm}
\bar{I}^{\pss (n)}_{\mu\nu}(k,k',k'')\;=\;\sum_{\rho\sigma}\,
\Theta^{\pss (n)}_{\rho\mu}(k-k'')\,
\Theta^{\pss (n)}_{\sigma\nu}(k'+k'')\,\Omega_F^{\pss (n-1)}(2k)\,\times
\nonumber \\
&&\left[\,-\eps_{\rho\sigma}(k,k',k'')\,D_{\pss FP}(-k')\,
\omega_F^\dagger(-k')\:-\:\omega_F(k)\,D_{\pss FP}(k)\,
{\cal E}^{{\pss (2)}\,\dagger}_{\rho\sigma}(-k,-k',-k'')\right. \nonumber \\
&&+\:\omega_F(k)\,D_{\pss FP}(k)\,R^{\pss (1)\, FP}_\rho(k,-k'')\,
D_{\pss FP}(k'')\,{\cal E}^{{\pss (1)}\,\dagger}_\sigma(-k',-k'')
\nonumber \\
&&-\:\epf_\rho(k,-k'')\,D_{\pss FP}(k'')\,R^{\pss (1)\, FP}_\sigma(k'',k')
\,D_{\pss FP}(-k')\,\omega_F^\dagger(-k') \nonumber \\
&&\left. -\:\epf_\rho(k,-k'')\,D_{\pss FP}(k'')\,
{\cal E}^{{\pss (1)}\,\dagger}_\sigma(-k',-k'')\,\right]\,
\Omega_F^{{\pss (n-1)}\,\dagger}(-2k')\, .
\eea

\end{document}